\documentclass[fleqn,usenatbib]{mnras}

\usepackage{newtxtext,newtxmath}

\usepackage[T1]{fontenc}

\DeclareRobustCommand{\VAN}[3]{#2}
\let\VANthebibliography\thebibliography
\def\thebibliography{\DeclareRobustCommand{\VAN}[3]{##3}\VANthebibliography}

\usepackage{subcaption}
\usepackage{graphicx}	
\usepackage{amsmath}	

\title[Starshades as technosignatures]{Starshades as technosignatures in direct imaging phase curves: Application to the Habitable Worlds Observatory targets}

\author[C. I. Skoglund and A. J. Mustill]{
Claudia I. Skoglund,$^{1,2}$\thanks{E-mail: claudia.skoglund@astro.su.se (CIS)}
Alexander J. Mustill$^{2}$
\\
$^{1}$The Oskar Klein Centre, Department of Astronomy, Stockholm University, AlbaNova University Center, SE-106 91 Stockholm\, Sweden\\
$^{2}$Lund Observatory, Division of Astrophysics, Department of Physics, Lund University, Box 118, SE-221 00 Lund, Sweden\\
}

\date{Accepted XXX. Received YYY; in original form ZZZ}

\pubyear{\the\year{}}

\begin{document}
\label{firstpage}
\pagerange{\pageref{firstpage}--\pageref{lastpage}}
\maketitle

\begin{abstract}
A star's luminosity increases as it evolves along the Main Sequence (MS), which inevitably results in a higher surface temperature for planets in orbit around the star. Technologically advanced civilizations may tackle this issue by installing artificial structures -- \emph{starshades} -- which can reduce the radiation received by the planet. Starshades, if they exist, are potentially detectable with current or near-future technology. We have simulated phase curve signatures in direct imaging of hypothetical starshades in systems targeted by the upcoming Habitable Worlds Observatory (HWO), which will be tasked with searching for Earth-like exoplanets orbiting nearby stars. The starshade is assumed to be a circular, reflecting surface placed at the inner Lagrange point between the star and the planet. Our results show that the phase curve of a starshade has a distinct shape compared to that of a typical planet. The phase curve signature lies above the expected $1\sigma=10^{-11}$ single-visit precision in contrast ratio of the telescope for $70.8\%$ of the target stars for the expected inner working angle (IWA) of around 60 mas. If the IWA can be reduced to 45 mas, the percentage of stars above the $1\sigma$ limit increases to $96.7\%$. With a sufficiently small IWA, HWO should be able to detect anomalies in light curves caused by starshades or similar highly-reflective surfaces -- which could serve as key indicators for technologically advanced civilizations.
\end{abstract}

\begin{keywords}
extraterrestrial intelligence -- stars: evolution -- planets and satellites: detection
\end{keywords}



\section{Introduction}
A star increases in luminosity and temperature as it evolves on the main sequence (MS) \citep{Gough1981}. This causes the liquid water habitable zone (HZ) to gradually move away from the star, where the HZ is defined as the range of distances to the star at which a terrestrial planet with
a suitable atmosphere can sustain liquid water on its surface \citep{Huang1959, Hart1978, Kopparapu2014}.For solar-type stars, an Earth-like planet will eventually find itself interior to the habitable zone, as the star's luminosity evolution will move this zone beyond the planet's orbit\footnote{This assumes the planet’s orbital semi-major axis does not change with time (for MS stars, we do not expect any significant stellar mass loss or tidal orbital decay).}, and the planet will experience a runaway greenhouse effect \citep{Rampino1994,OMalley2013, Gaidos2017}. The motion of the HZ away from stars during their evolution means that any technological civilisation will need to maintain the habitability of orbiting planets. As discussed by \cite{Gaidos2017}, possible solutions include altering the stellar output, actively changing the climate of the planet \citep[examples of this are discussed by][to mitigate sea-level rise]{Minunno2023}, or deflecting some of the radiation away from the planet. The latter of these may be preferable for a number of reasons: it is more predictable than actively changing the climate of the entire planet and also less technically challenging than altering the stellar output \citep{Gaidos2017}. By introducing a highly reflective surface (hereafter called a `starshade') placed between the star and the planet, the flux received by the planet can be reduced in order to ensure its habitability. 

An obvious choice for placement of a starshade is the inner Lagrange point, $\mathcal{L}_1$, between the star and the planet. Not only does this allow the blocking out of light from the star, but it also minimizes the need for station-keeping. Both the inner and outer Lagrange points of Earth are already used as parking spots for spacecraft exploring the universe \citep[see e.g.][]{Sugiura2023}.

Reflecting or refracting surfaces placed between the Sun and the Earth have been proposed in several other studies in the context of anthropogenic warming \citep[e.g.][]{Govindsamy2000, Angel2006, Sanchez2015, Fuglesang2021}. However, while climate change of industrial origin may not be a general problem for all potential extraterrestrial civilizations, the increasing luminosities of evolving stars will be, should the civilisation live long enough. Starshades may therefore be potential technosignatures in the search for extraterrestrial intelligence, or SETI \citep{Gaidos2017}.  As opposed to biosignatures, which are atmospheric or surface-originating spectral features serving as evidence for extraterrestrial life of any complexity \citep[e.g.][]{Schwieterman2018, Madhusudhan2019, Wordsworth2022}, technosignatures could directly point toward a technologically advanced civilisation and might even be more detectable, abundant, long-lived, and unambiguous than the biological counterpart \citep{Wright2022}. 

The detection of exoplanets includes various techniques, such as the transit method, radial velocity, direct imaging, and gravitational microlensing. Technosignatures could appear in the data resulting from those methods in various forms \citep[see e.g.][]{Amiri2024, Kayali2024, Tusay2024, Tremblay2025}. The increasing interest in potential technosignatures shows that it is necessary to rigorously assess them and establish well-defined observational expectations, ensuring their recognition if they were to manifest in astronomical data. In the case of starshades, it has already been shown that they may appear in detection as unusual transit light curves \citep{Gaidos2017}. Considering previous efforts of evaluating technosignatures in terms of the expected capabilities of the Habitable Worlds Observatory (HWO) \citep[see][]{Kopparapu2024, Dubey2025}, we will investigate starshades in the context of direct imaging. For starshades, and highly reflective surfaces in general, we may anticipate that the light received by an observer in direct imaging will be different compared to a planet without such a structure.

Direct imaging with current or near-future technology does not, however, provide us with fully resolved views of exoplanets. For a directly imaged planet, the radius and albedo are degenerate parameters and can only be very loosely constrained, unless one of them is determined by a different technique. Instead, at a given star--planet--observer phase angle, the relevant measure is the planet-to-star flux ratio which is wavelength dependent \citep{Salvador2024}. From this, we can construct direct imaging phase curves, which show the varying brightness of a planet as it orbits its host star \citep{Parmentier2018}. Phase curves can offer valuable insights into atmospheric composition, surface properties and temperature variation. In direct imaging phase curves, the flux ratio of the planet--star system can be anticipated to increase when introducing a highly reflective starshade. However, given the limited constraints on the planet's radius and albedo, this would not be enough information to provide a clear technosignature, if only a few epochs of observations were made. More importantly, placing the starshade at $\mathcal{L}_1$ will give a unique geometry that makes the observed reflected light different depending on where the planet is in its orbit; while the planet is a sphere, a simple shade would be a flat circular disc and will therefore behave differently compared to the planet as seen from different angles. We note that the distance between $\mathcal{L}_1$ and the planet is expected to be too small to resolve the starshade and planet from each other (see Section \ref{sec:starshade_placement}), and the phase curve will carry information about their combined reflected light.

The technosignatures expected to be detectable from phase curves are, nevertheless, limited by the capacity of today's technology. Over the last decade, direct imaging efforts have been improved at a promising pace. The method has been used to discover several giant planets with long orbital periods \citep[e.g.][]{Chauvin2004, Marois2008, Lagrange2009}. However, currently direct imaging detections have been limited to such planets. The successful detections typically have a planet-to-star contrast ratio of about $10^{-4}$ to $10^{-5}$, in visible light, while an Earth analogue would have a ratio of $10^{-10}$ \citep{Li2021}. As such, current telescopes do not allow for direct imaging of Earth-sized planets \citep{HABEX2020}. Currently under production and previously called the Wide Field Infrared Survey Telescope \citep[WFIRST;][]{Spergel2015,Akeson2019}, the Nancy Grace Roman Space Telescope\footnote{Named after former NASA Chief of Astronomy Nancy Grace Roman who proposed detection of exoplanets by space-based direct imaging as early as 1959 \citep{Roman1959}. The authors owe this observation to Dominic Benford, via a conference talk by Beth Biller at Exoplanets 5, Leiden, 2024.} holds great promise in detecting smaller contrast ratios. Still, Earth-like planets with low planet-to-star flux ratios will go undetected. The subsequent generation of telescopes might, however, be able to solve this problem. A revolutionary performance in terms of the lower limit for planet-to-star flux ratios is proposed with the Habitable Worlds Observatory (HWO) developed by NASA. HWO builds upon earlier suggested mission concepts such as the Habitable Exoplanet Observatory (HabEx) \citep{HABEX2020} and the Large Ultraviolet Optical/Infrared Surveyor (LUVOIR) \citep{Peterson2017, Roberge2018, Luvoir2019}. The proposed 6-m telescope is expected to achieve significant detections at contrast ratios below $10^{-10}$ \citep{Decadal2020}. Thus, the concept is expected to allow direct imaging of exo-Earths. The telescope is planned to target nearby stars ($d<25$\,pc), and a target list was compiled by \cite{Mamajek2024}, motivated by the Astro 2020 Decadal Survey. These stars have been selected as they have HZs far enough from the star for the telescope to get a clear view of potentially habitable planets \citep[see][]{Mamajek2024}, and any Earth-sized planet located in the HZ would be bright enough to allow spectral analysis of its atmosphere.

In this paper we apply the starshade concept to these prospective HWO targets. We generate phase curve signatures for starshades orbiting the stars that meet our conditions in terms of stellar age and planet--star separation. We thereby  demonstrate the detectability of such structures with near-future technology. 

The paper is structured as follows. In Section \ref{sec:math_mod_form}, we describe the stellar catalogue as well as the mathematical model that underlies our simulations. The results from the simulations built on this model are then presented in Section \ref{sec:results}. We provide a discussion of these results in Section \ref{sec:disc} and a brief summary of our work as well as conclusions in Section \ref{sec:conclusions}.

\section{Mathematical Model Formulation}
\label{sec:math_mod_form}

\subsection{Methodology and underlying assumptions}
\label{sec:Method}
We assume that each HWO target star, with current age $t_\mathrm{age}$, hosts a technological civilisation that seeks to stabilise the surface temperature at the value it was at some target age $t$ in the past, where $t=[1,3,5,7,9]$ Gyr as measured from the formation of the system. We then calculate the starshade size at $t_\mathrm{age}$ for each of the target ages $t<t_\mathrm{age}$. The stellar age is obtained by isochrone fitting using the current luminosity $L_*$, effective temperature $T_\mathrm{eff}$ and metallicity of each star in the HWO target list provided by \cite{Harada2024} (see Section \ref{sec:stellar_evo_iso}). We then calculate the resulting phase curve signature for a range of values of orbital inclination, and determine whether the starshade's contribution will be detectable. 

The ages $t$ have been chosen on the assumption that intelligent life takes a long time to develop. Recent estimates suggest that life on Earth began more than 3.5 Gyr ago, with some indications it could have emerged as early as 4.5 Gyr ago \citep[see, e.g.][]{Pearce2018, Prosdocimi2023}. Moreover, in order for each target age $t$ to be considered, the star must be 0.5 Gyr older than that age. This ensures that the size of the starshade is large enough at the current age to give a significant technosignature. Based on these assumptions, stars below the age of 1.5 Gyr, 44 in total, have been removed from the sample.

At each target age, we assume that the star has an Earth-sized planet at the Earth Equivalent Distance (EED). The EED is defined as the distance to a star at which a hypothetical planet would receive the same amount of radiation from its host star as Earth currently does from the Sun. 

For the starshade size calculations and phase curve simulations, we assume that the starshade is a perfectly flat, circular disc that exhibits Lambertian reflection and is placed at the inner Lagrange point $\mathcal{L}_1$. Moreover, we assume the geometric albedo to be 0.4 for the planet and 0.9 for the starshade. The geometric albedo of the starshade can be considered an optimistic, highly-reflective case (see Section \ref{sec:rep_case} for a discussion of the difference the albedo makes). The geometric albedo of a planet varies with wavelength; our adopted value corresponds to the Earth's R-band geometric albedo of 0.418 \citep{Mallama2017}. This is an approximation as we do not specify a particular wavelength range. While the actual inner working angle (IWA) of the HWO mission is currently under discussion \citep[see e.g.][]{Vaughan2023}, for HabEx the expected IWA was around 62 mas in the visible \citep{HABEX2020}. However, e.g. \cite{Vaughan2023} proposes a smaller IWA in order to access more scattering features of the targeted planets. We have consequently chosen to apply an IWA of both 60 mas and 45 mas in our phase curve simulations.

\subsection{Stellar evolution and isochrone fitting}

\label{sec:stellar_evo_iso}

\begin{figure}
\centering
	\includegraphics[scale=0.55]{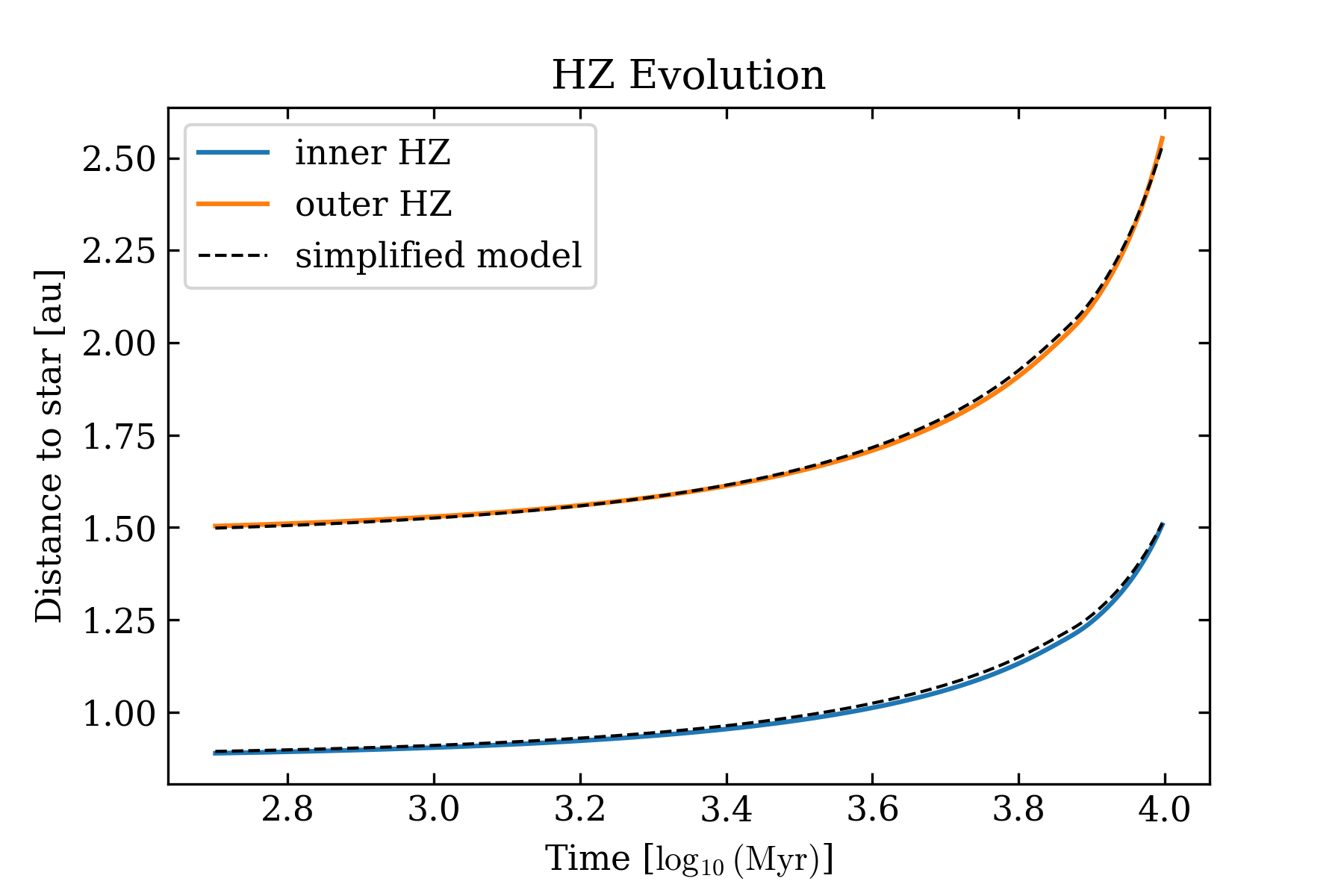} 
    \caption{A visualization of the HZ distances around a sun-like star as a function of stellar age. The solid lines are based on the HZ model of \protect\cite{Kopparapu2014} and the dashed lines correspond to our simplified model assuming no dependence on the stellar effective temperature. }
    \label{fig:HZ_evol}
\end{figure}

Both stellar luminosity and radius increase during the star's MS lifetime, and this changes the distance of the HZ to the star; an example is shown in Fig.~\ref{fig:HZ_evol}. These distances typically depend on both the luminosity $L_*$ and effective temperature $T_\mathrm{eff}$ of the star \citep[see e.g.][]{Kopparapu2014}. However, the temperature dependence is typically below $10\%$ and for our purposes this is negligible compared to the dependence on the luminosity  (the HZ distances goes as $\sqrt{L_*/L_\odot}$). Therefore, we have chosen to model the starshade size solely on the luminosity evolution of the star. We also consider the increasing stellar radius when modelling the starshade size, as it affects the geometry of the system and how much of the stellar disc is covered by the starshade (see below). 

\begin{figure}
\centering
	\includegraphics[scale=0.55]{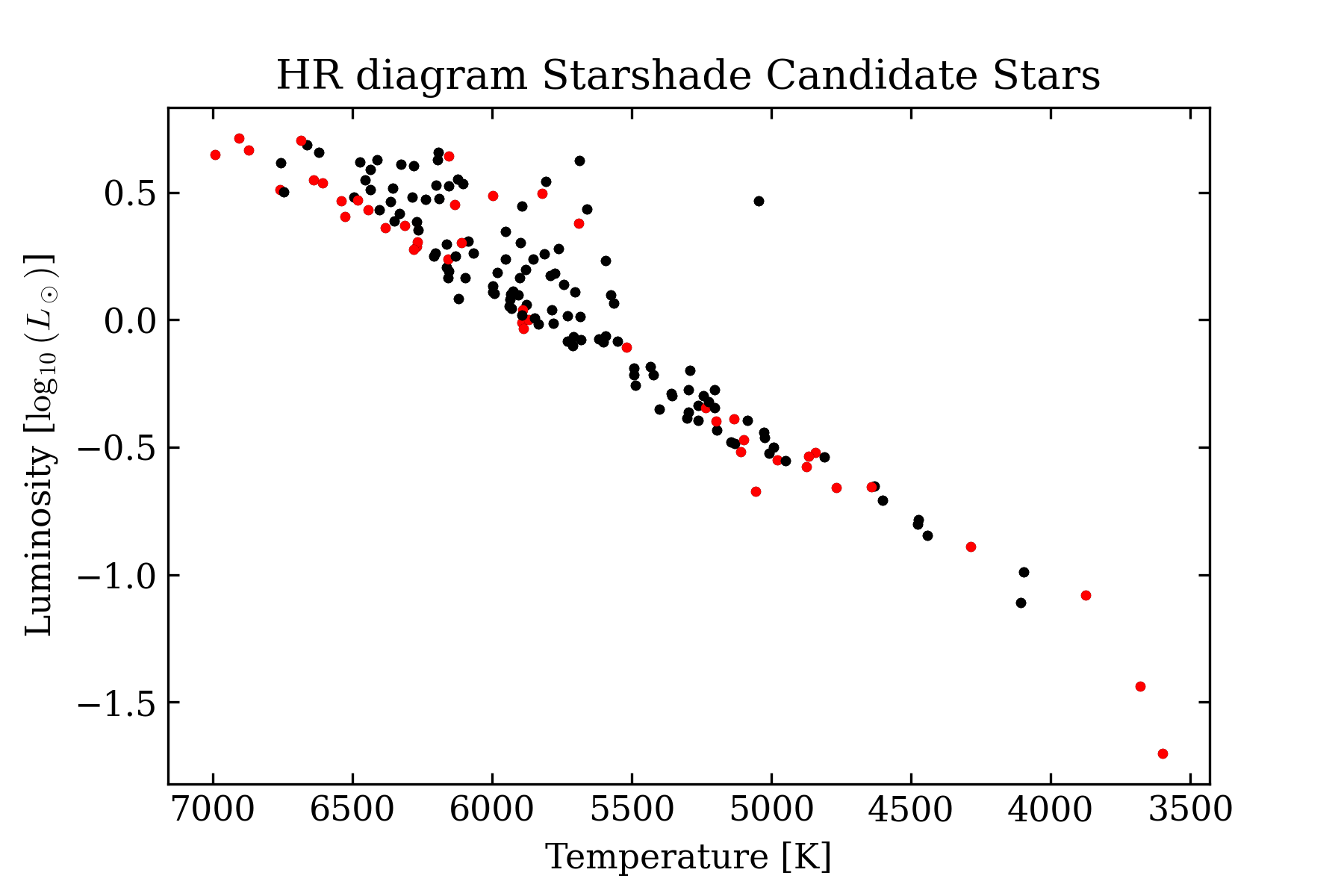}
    \caption{HR diagram of the HWO target stars. The stars below 1.5 Gyr according to our isochrone fitting are shown as red dots and will not be considered as potential starshade hosts. }
    \label{fig:HR_plot}
\end{figure}

\begin{figure}
\centering
	\includegraphics[scale=0.55]{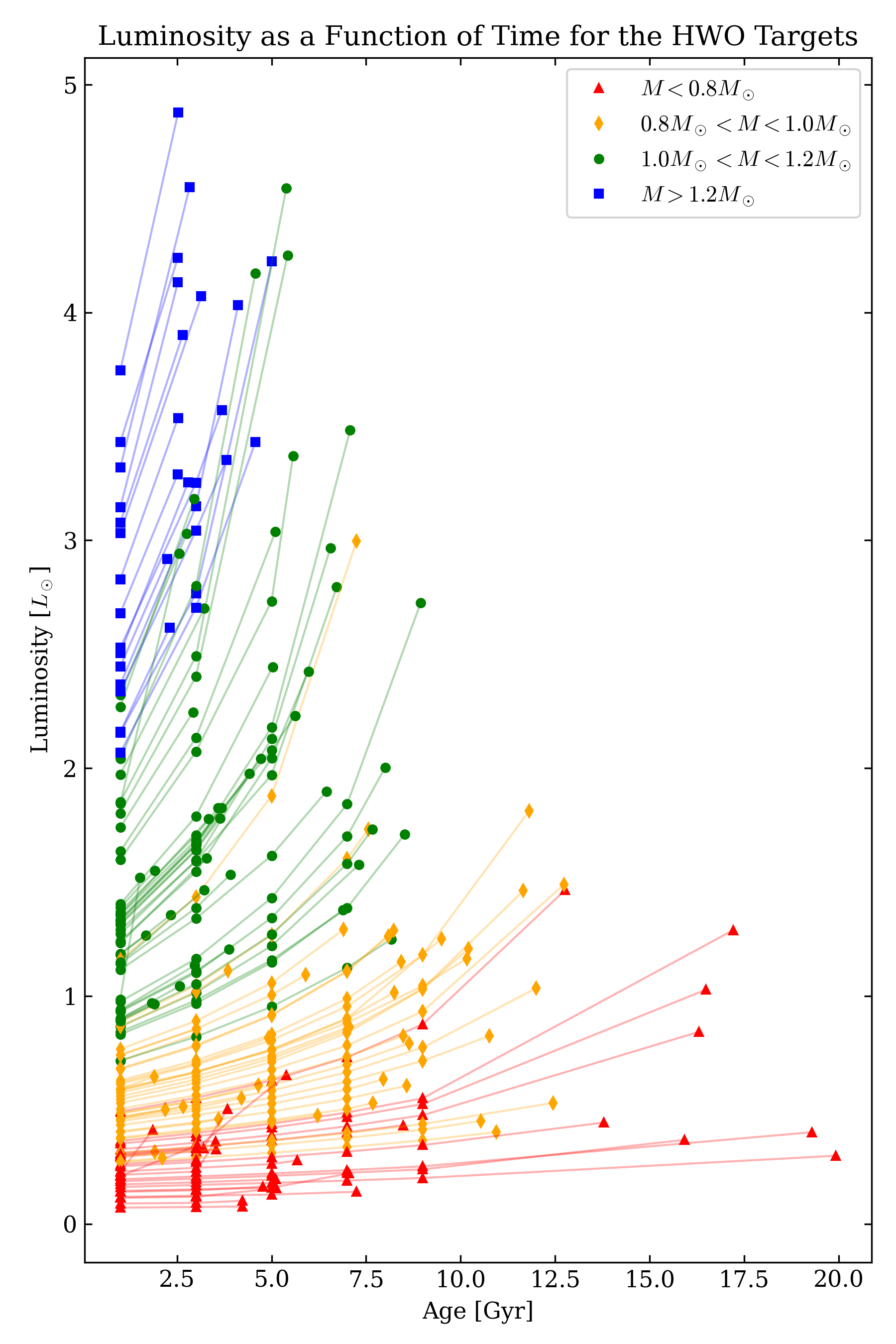}
    \caption{The luminosity as a function of time for the starshade candidate stars. The dots represents the data points, which correspond to the ages $t= [1,3,5,7,9]$ Gyr as well as the current age $t_\mathrm{age}$. The luminosity was found for each time point using MIST isochrones.}
    \label{fig:luminosity_time_HWO}
\end{figure}

To homogeneously derive stellar ages and masses, we performed isochrone fitting on all HWO target stars in a version of the HWO target list by \cite{Harada2024}. Updated stellar parameters to that in \cite{Mamajek2024} are found in both in \cite{Harada2024} and \cite{Tuchow2024}. The list is dominated by MS dwarfs, and contains only a few subgiants, and no giants. Out of the 164 stars, 66 are spectral type F, 55 type G, 40 type K and 3 type M; an HR diagram is shown in Fig~\ref{fig:HR_plot}.
Isochrones were downloaded from MESA Isochrones \& Stellar Tracks (MIST) \citep{Dotter2016, Choi2016, Paxton2011, Paxton2013, Paxton2015} using \textsc{ezmist}\footnote{Morgan Fouesneau \& contributors: \url{https://github.com/mfouesneau/ezmist}} for all different metallicities presented in the HWO target list by \cite{Harada2024}. Based on the stellar luminosity $L_*$ and effective temperature $T_\mathrm{eff}$ in the HWO target list, the corresponding mass ($M_*$), age ($t_\mathrm{age}$) and radius ($R_*$) were fitted (each against $L_*$ and $T_\mathrm{eff}$) by interpolation. The stars younger than $1.5$ Gyr were then filtered out as seen in Fig.~\ref{fig:HR_plot}. All of the stars in the HWO sample are around solar mass and consequently the mass loss on the MS is negligible. Alongside the values at the current age for each star, the corresponding $L_*$ was also found for each target age $t=[1,3,5,7,9]$ Gyr using the same method for isochrone fitting, allowing for calculation of the Earth Equivalent Distance (EED) at the target ages. The luminosity as a function of time (calculated for our grid of target ages) for all HWO targets above 1.5 Gyr is shown in Fig.~\ref{fig:luminosity_time_HWO}. 

\subsection{Starshade placement}
\label{sec:starshade_placement}

The inner Lagrange point $\mathcal{L}_1$ has been suggested as a point of placement for a potential geoengineering solution to anthropogenic climate change \citep{Govindsamy2000, Angel2006, Sanchez2015, Szapudi2023} as well as to warming due to the evolution of the Sun \citep{Gaidos2017}. The placement at $\mathcal{L}_1$ assumes that it would be possible with future technology to keep the structure stable for a long period of time at that specific point. Moreover, the choice of placement at precisely $\mathcal{L}_1$ assumes that the radiative force does not contribute by a significant amount, which essentially means that the starshade needs to have relatively large mass. In \cite{Gaidos2017}, the starshade mass was minimized and hence the radiative force was taken into consideration. This moves the placement of the starshade to some distance away from $\mathcal{L}_1$, closer to the star. Even if this can mean that a smaller amount of material is needed for construction, we argue that a more massive starshade also has its advantages since it is more robust and hence can better withstand collisions with objects from space. We choose to place the starshades at $\mathcal{L}_1$ exactly.

The fractional distance $d_{\mathrm{L}1}$ (the distance between the smaller body and $\mathcal{L}_1$ divided by the distance between the smaller and larger body) can be evaluated via 
\begin{equation}
    \mu_1 \left( - \frac{1}{(1-{d_{\mathrm{L}1})}^2}+1-d_{\mathrm{L}1} \right)-\mu_2\left(- \frac{1}{d_{\mathrm{L}1}^2} + d_{\mathrm{L}1}\right)=0
    \label{eqn:d_l1}
\end{equation}
where $\mu_1=GM_1$ and $\mu_2=GM_2$ are the gravitational parameters of the larger and smaller body respectively \citep{Murray2000}. Moreover, $G$ is the gravitational constant, $M_1$ is the mass of the larger body and $M_2$ that of the smaller body. For context, the fractional distance for the Sun--Earth system is $d_{\mathrm{L}1}\approx 0.01$. This will roughly be the case for all our targets as we are assuming an Earth-mass planet and the stars are all around solar mass. 

We expect that the distance from the planet to $\mathcal{L}_1$  for all stars on the target list will be unresolvable with the HWO: the angular separation in our sample lies below $\sim 7$ mas, below HWO's likely diffraction limit, which would be $\sim20$\,mas at $0.6\,\mu$m.

\subsection{Starshade size}
\label{sec:starshade_size}
The size of the starshade will depend on the amount of starlight that must be blocked out to maintain a constant flux $F_\mathrm{want}$ on the surface of the planet as the star ages. \textbf{$F_\mathrm{want}$} was set to be the current solar flux on Earth. This means that the hypothetical exoplanets were set to be at the EED, which varies depending on age and mass of the star at the target age in the past. Naively, one could express the desired flux as
\begin{equation}
    F_\text{want}=F_\text{new} \left(1-\frac{\mathcal{A}_\text{block}}{\mathcal{A}_*} \right).
    \label{eqn:naivearea}
\end{equation}
Here, $F_\text{want}$ is the flux that is to be kept constant by blocking out a certain angular area $ \mathcal{A}_\text{block}$ of the stellar disc with angular area $\mathcal{A}_*$ when the increased flux is equal to $F_\text{new}$.

\begin{figure*}
\centering
	\includegraphics[scale=0.9]{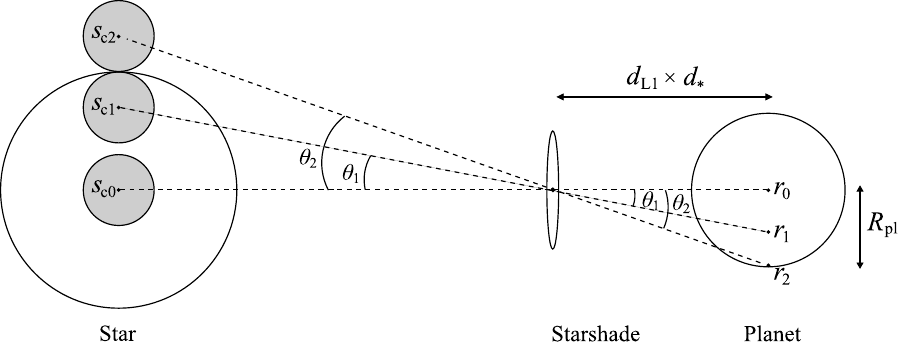}
    \caption{Visualization of where the starshade will be placed on the stellar disc seen from
different points on the planet. The points $s_{\mathrm{c}i}$ are the points on which the starshade appear to be centred as seen from the points $r_i$ on the planet disc. The points $r_1$ and $r_2$ in particular denote when the starshade starts to leave the stellar disc and when the starshade completely leaves the stellar disc respectively. In the integrals found in equation (\ref{eqn:integrals}), these two represent the distances with respect to the centre of the planet disc. (Not to scale.)}
    \label{fig:angles}
\end{figure*}

This calculation of $\mathcal{A}_\text{block}$ assumes that the starshade blocks out an equal amount of light seen from all points on the planet. When observing the starshade from the planet's substellar point, the starshade appears to be centred on the stellar disc. This is not the case when observing it from other points, such as from the North Pole or South Pole. Fig.~\ref{fig:angles} visualizes this. In the figure, $S_{\mathrm{c}i}$ denotes the different points on which the starshade will appear to be centred seen from different points on the planet. Thus, different amounts of stellar flux reach different parts of the planet. For Earth specifically, we found that the starshade will not be completely inside the solar disc when observing it from the North Pole or South Pole (i.e. at high/low latitudes). Therefore, we improved on equation~(\ref{eqn:naivearea}) to account for such cases. 

Given the geometries in Fig.~\ref{fig:angles}, one can estimate the total power received by the planet by integrating the flux over the whole planetary disc. Hence, we get: 
\begin{equation}
\begin{aligned}
    P &= \int_0^{r_1} 2\pi r F_0 \left( 1 - \frac{\mathcal{A}_\text{block}}{\mathcal{A}_*}  \right) \, \mathrm{d}r \\
    &\quad + \int_{r_1}^{r_2} 2\pi r F_0 \left( 1 - \frac{\mathcal{A}(r)}{\mathcal{A}_*}  \right) \, \mathrm{d}r \\
    &\quad + \int_{r_2}^{R_\mathrm{p}} 2\pi r F_0 \, \mathrm{d}r.
\end{aligned}
\label{eqn:integrals}
\end{equation}
Here, $F_0$ is the flux density without the starshade and $\mathcal{A}(r)$ is the blocked out area as the starshade leaves the stellar disc, as seen from a point at a projected radius $r$ from the centre of the planet. After the point at which the starshades leaves the stellar disc, $\mathcal{A}(r)$ becomes increasingly smaller as we move further out from the centre of the planet. It can be evaluated from typical mathematical formulas for circle-circle intersections \citep[see e.g.][]{weisstein_circle_intersection}. The radius of the planet is denoted by $R_\mathrm{p}$. $r_1$ is the point on the planet for which the starshade starts to leave the stellar disc and $r_2$ is the point for which it has left the disc completely (see Fig.~\ref{fig:angles}). This means that, if the starshade does not leave the stellar disc at all, only the first term is needed and the desired flux reduces to that of equation~(\ref{eqn:naivearea}) (while the second and third terms are added for the cases when it leaves the disc partly and completely respectively). In our sample, the starshade left the stellar disc, to some extent, for 97 of the stars.

Equation \ref{eqn:integrals} allows for a forward-modeling of the flux incident on the planet for a given starshade size. Instead, we need to calculate the starshade size given the flux, and equation (\ref{eqn:integrals}) is not invertible analytically. Therefore, inversion was performed by the bisection method.

\subsection{Phase curves}
\label{sec:phase_curves}
The phase curve is the variation in reflected planetary light received by the observer as the planet orbits its star. It is set by both the change in visible illuminated area as the planet orbits and the efficiency with which the planet and its atmosphere scatter light at different angles. The phase curve itself is constructed from a phase function, which depends on the phase angle $\alpha$ (the star--planet--observer angle). Since the planet reflects different amounts of light depending on where it is in its orbit, the phase curve will have a characteristic shape depending on the amount of reflected light at different points. The phase angle, which determines the observable amount of light that a planet reflects, can be calculated from
\begin{equation}
    \cos \alpha = \sin(\omega+\nu) \sin i
    \label{eqn:phaseangle1}
\end{equation}
where $\omega$ is the argument of periapsis, $\nu$ is the true anomaly and $i$ is the orbital inclination \citep[e.g.][]{Kane_2010, Madhusudhan2012, Bruna2023}. For circular orbits, this simplifies to $\omega + \nu= \theta$ where $\theta$ is the orbital longitude measured from the line of nodes. This means that equation (\ref{eqn:phaseangle1}) becomes
\begin{equation}
    \cos \alpha = \sin \theta \sin i.
\end{equation}

\begin{figure*}
\centering
	\includegraphics[scale=0.55]{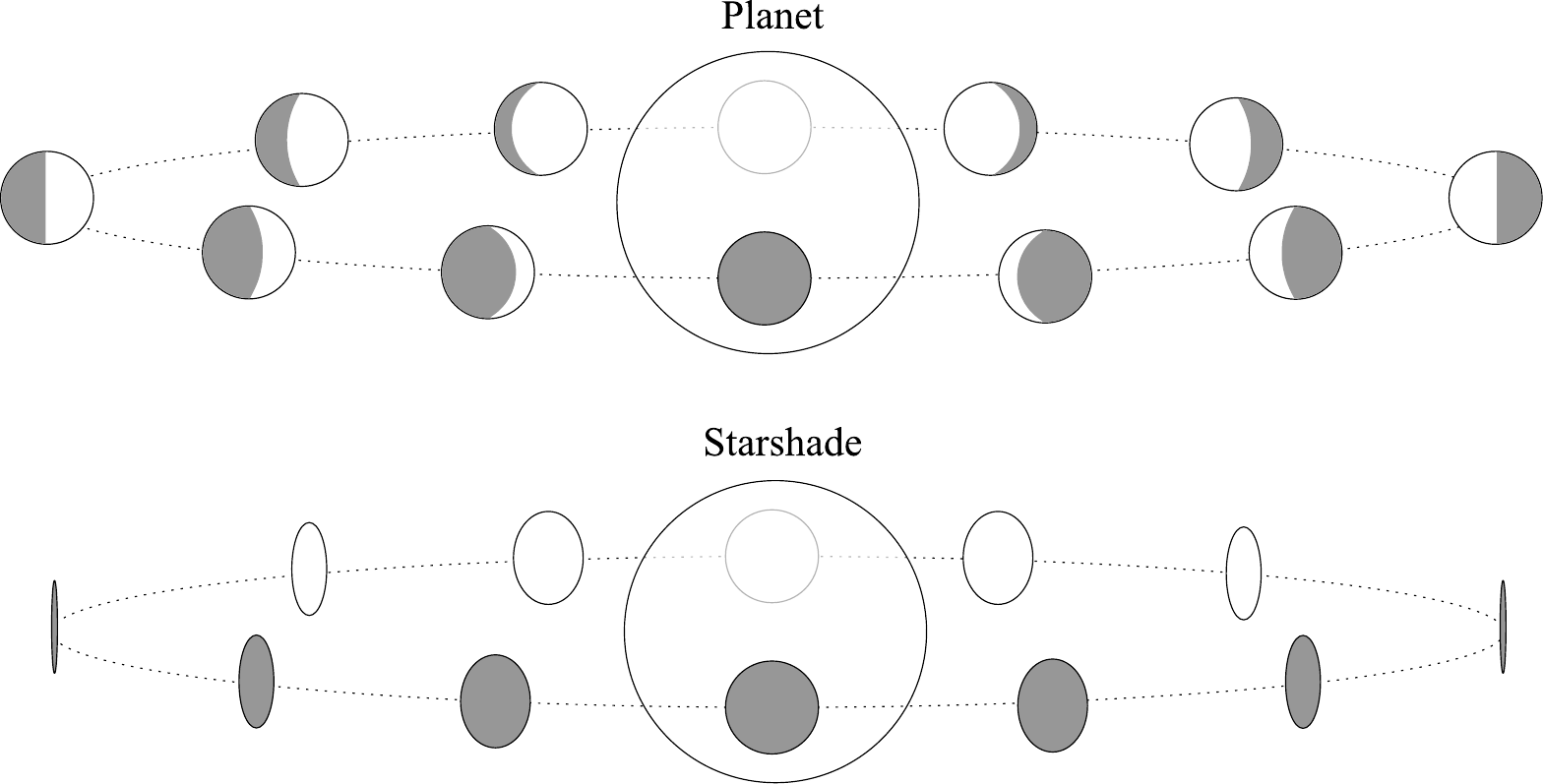}
    \caption{Visualisation of the illuminated surface of a planet (top) and of a circular “disc”
starshade (bottom). (Not to scale.)}
    \label{fig:phases}
\end{figure*}

Placing a reflective starshade at $\mathcal{L}_1$ will yield a different amount of reflected light when observing the planetary system compared to a planet by itself. For a discussion on how other structures, such as planetary rings and clouds, affect the phase curve we refer to Section \ref{sec:detectability} We show a visualisation of how the starshade reflects light differently compared to a spherical planet in Fig.~\ref{fig:phases}. Since the starshade is a circular disc, it will not generate any extra reflected light if the planetary system is observed face on. With some inclination, however, the excess light reflected by the starshade may yield a phase curve distinguishable from that of only the star and the planet. 

\begin{figure}
\centering
	\includegraphics[scale=3]{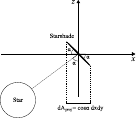}
    \caption{The geometry for the derivation of the starshade phase function. Here, the y-axis points into the paper, and the observer looks down the z-axis. (Not to scale.)}
    \label{fig:starshade_phase}
\end{figure}

The hypothetical starshade may exhibit different scattering mechanisms depending on its engineering properties. However, since the scope of possible designs for such structure is immense, it seems unreasonable to complicate the equations with different scattering mechanisms. Hence, we assume simple Lambertian reflection for the starshade. We aim to show that this simple model gives rise to a noticeably different phase curve compared to a planet alone -- we may even get stronger signals if the structure is metallic and exhibits specular reflection. Given the geometry in Fig.~\ref{fig:starshade_phase}, the phase function for the starshade can be expressed as
\begin{equation}
    \Psi_\mathrm{shade}= \cos\alpha.
    \label{shade_phasefunc}
\end{equation}
The phase function in equation~(\ref{shade_phasefunc}) only holds when $\cos \alpha \geq 0$. When $\cos \alpha <0$, the starshade will be facing away from the observer, and assuming negligible forward-scattering and that its rear side is not emitting, this means that the phase function will be equal to zero. 

The phase functions of planets are, none the less, more widely studied. We have employed a phase function model by \citet{Heng2021} in our calculations, assuming a single-scattering albedo of $\omega=0.9$. For the mathematical formulation of this function, encapsulating several different scattering mechanisms, we refer the reader to \cite{Heng2021}. Using this model allows us to account for e.g. Rayleigh scattering, which is expected for planets with Earth-like atmospheres. Compared to a Lambertian sphere, this gives a slightly more peaked phase function around opposition. The flux ratio of the planet and the star is then given by
\begin{equation}
    \frac{F_\mathrm{p}}{F_*}=\left(\frac{R_\mathrm{p}}{a_\mathrm{p}} \right)^2 A_\mathrm{p} \Psi_\mathrm{p}
\end{equation}
for which $\Psi_\mathrm{p}$ is the phase function of the planet as defined in \cite{Heng2021}, $R_\mathrm{p}$ is the radius of the planet and $a_\mathrm{p}$ is the orbital semimajor axis. $A_\mathrm{p}$ is the planet's geometric albedo, which is defined as the ratio of the brightness seen at zero phase angle to that of an idealized, fully reflective Lambertian disc with the same cross section as the planet \citep{Seager2010}. Similarly, the flux ratio for the starshade--star system can be expressed in terms of the starshade phase function stated in equation (\ref{shade_phasefunc}):
\begin{equation}
    \frac{F_\mathrm{shade}}{F_*}= \left(\frac{R_\mathrm{shade}}{a_\mathrm{shade}} \frac{}{}\right)^2 A_\mathrm{shade} \Psi_\mathrm{shade}.
\end{equation}

Even though the flux ratio formula can be obtained for the shade and the planet separately, simply adding the two together is not enough to describe the phase curve of the system. This is because the starshade is being placed in front of the planet, which reduces the flux the planet receives and hence also how much it reflects. As the planet moves around in its orbit, it will reflect different amounts of light to the observer depending on how the shadow falls on the planet. To account for this, we consider a simplified model where we average out the received power by the planet and incorporate it as a factor in the flux ratio. In this way, we account for the planet being sited in the shade's penumbral shadow \citep{Baum2022}, which we consider to be independent of position on the planet. This is not strictly true when the planet is seen around quadrature, especially for the extreme case of nearly face-on inclination, where the observer mostly sees the region round the planet's terminator. The difference this makes will be small, since the starshade never fully leaves the stellar disc for any of our targets: in the most extreme case in our sample, 70\% of the starshade is still obscuring the stellar disc when viewed from the terminator. This means that the gradient of the shadow from the substellar point to the terminator is not so extreme. Consequently, the flux ratio for the shadowed (i.e. reduced flux) planet becomes
\begin{equation}
    \frac{F_\mathrm{p,red}}{F_*}=\left( \frac{R_p}{a_p} \right)^2 A_g \Psi_\mathrm{p} P_\%.
\end{equation}
The factor $P_\%$ is the power from the stellar radiation that the planet actually receives with a starshade included in the system [equal to $P$ in equation~(\ref{eqn:integrals})] divided by the power the planet would receive without the starshade (which simply depends on the luminosity of the star and the planet--star distance). 

Now, the flux ratio of the shaded planet can be added to that of the starshade to obtain the total flux ratio of the system: 
\begin{equation}
    \frac{F_\mathrm{tot}}{F_*}=\frac{F_\mathrm{p,red}}{F_*}+\frac{F_\mathrm{shade}}{F_*}.
\end{equation}
The flux ratio can be plotted against the orbital longitude, which gives us the phase curve.

Depending on the IWA of the telescope, part of the calculated phase curve will not be visible to the observer. We assume no meaningful data can be collected when the planet is inside the IWA. Assuming a circular orbit and using the small angle approximation, the on-sky projected distance $r_\mathrm{proj}$ of the planet can be expressed as 
\begin{equation}
    \sin(\alpha)=\frac{r_\mathrm{proj}}{a}
\end{equation}
\citep{Vaughan2023}, where $a$ is the semimajor axis and $\alpha$ is the phase angle depending on orbital inclination and longitude as previously defined. From the projected distance $r_\mathrm{proj}$ as well as the distance from the observer to the star, $D_*$, we calculate the on-sky angular separation $\delta$;
\begin{equation}
    \delta=\frac{r_\mathrm{proj}}{D_*}.
    \label{eqn:ang_sep}
\end{equation}
This means that when $\delta<$ IWA, the planet will not be visible in the phase curve and we consequently remove these parts in our simulations. For HWO in particular, we assume an inner working angle of 60 mas building upon the HabEx concept \citep{HABEX2020} (where an IWA of 62 mas is proposed) as well as a lower IWA of 45 mas.

\section{Results}

\label{sec:results}

\subsection{Representative cases}

\label{sec:rep_case}

We now show some illustrative examples of the required starshade size as a function of stellar age, as well as the resulting phase curves. 

\begin{figure}
    \begin{subfigure}{.49\textwidth}
        \centering
        \includegraphics[scale=0.55]{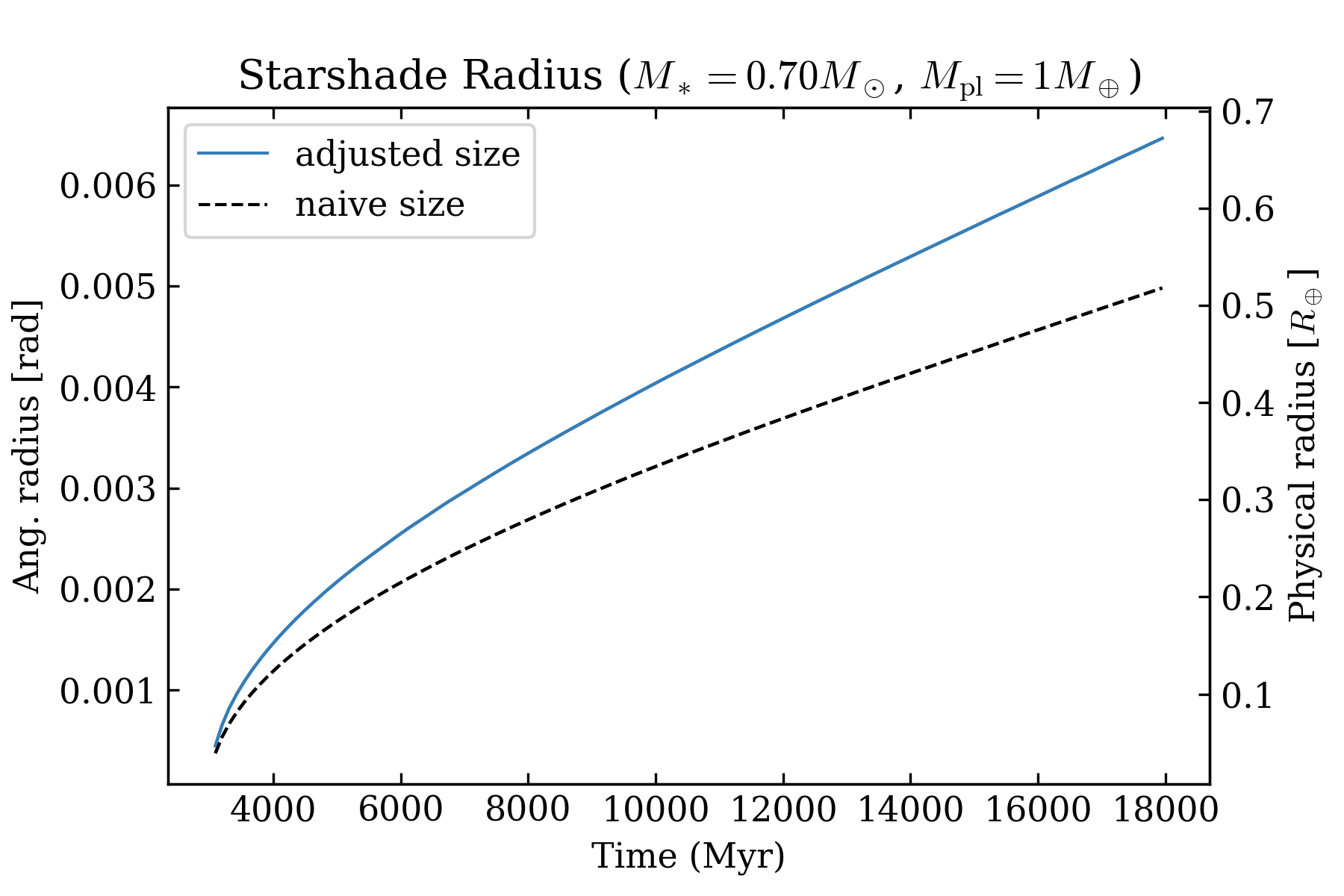}
  
        \caption{}
    \end{subfigure}\hfill
    \begin{subfigure}{.49\textwidth}
        \centering
        \includegraphics[scale=0.55]{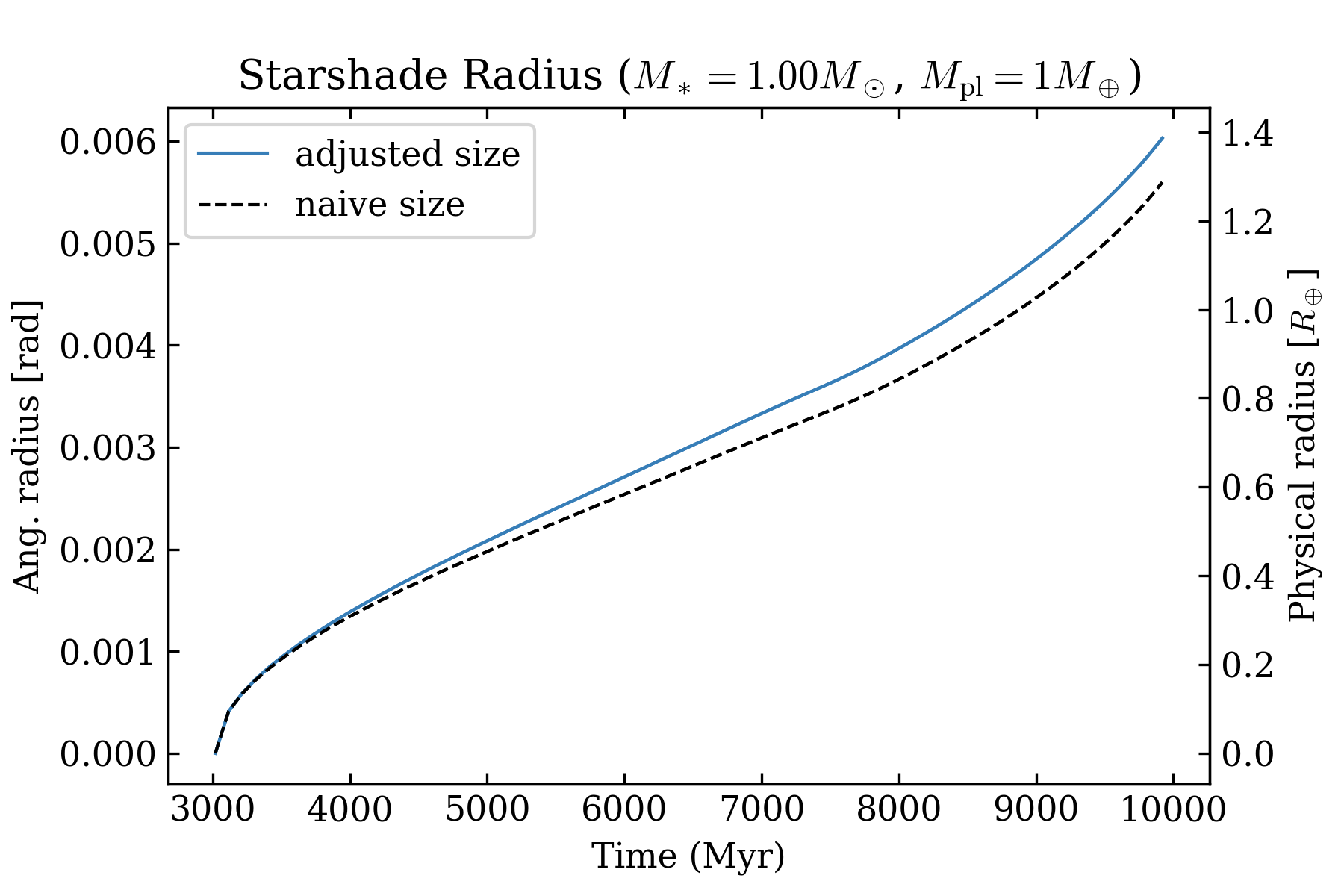}
     
        \caption{}
    \end{subfigure}
        \begin{subfigure}{.49\textwidth}
        \centering
        \includegraphics[scale=0.55]{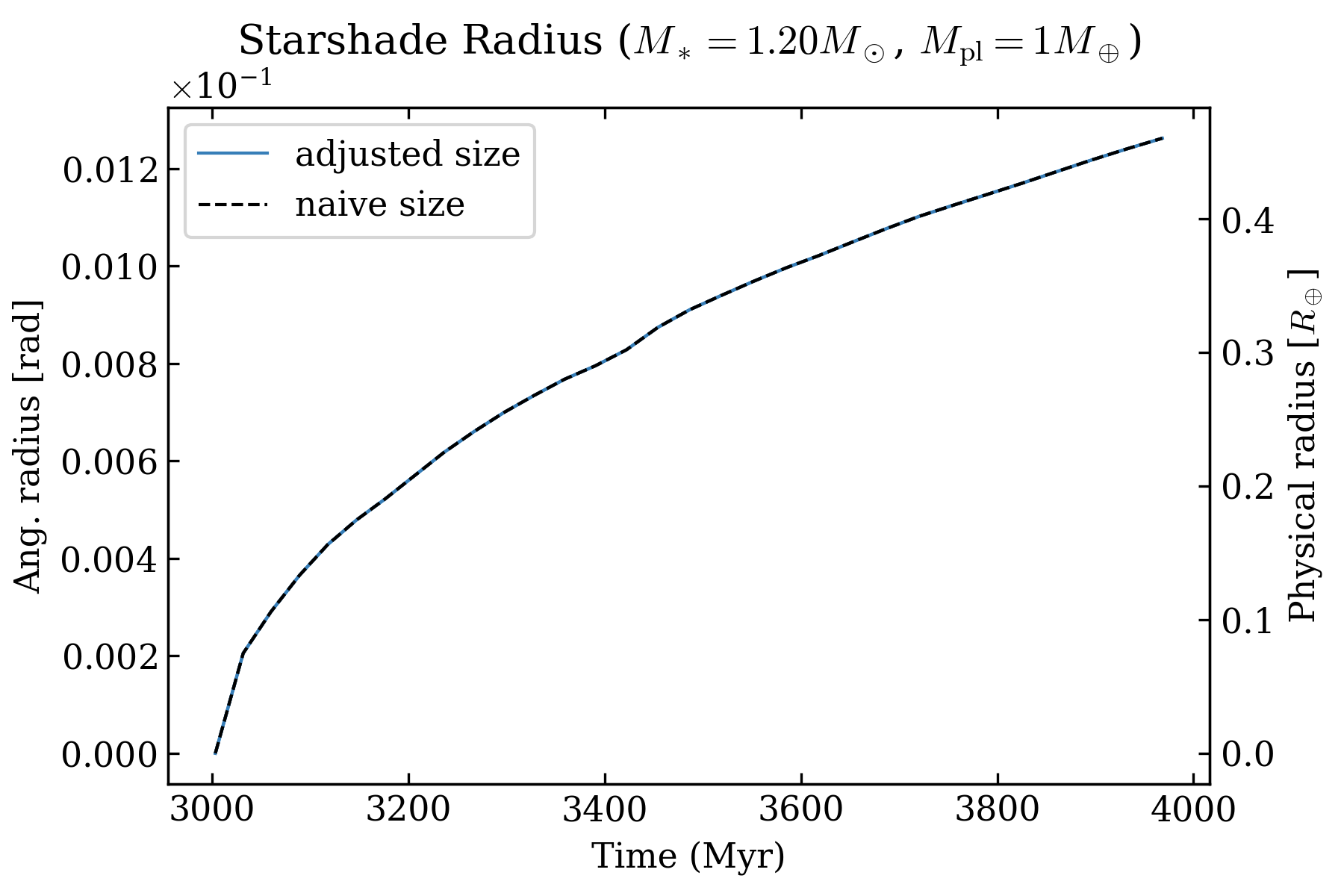}
    
        \caption{}
    \end{subfigure}
    \caption{Starshade size as a function of time for a solar metallicity star with a $1 M_\oplus$ planet placed at the Earth Equivalent distance at 3 Gyr. The size changes so that the flux received by the planet stays equal to that at 3 Gyr. The different panels represent three different stellar masses: (a) $0.7 M_\odot$, (b) $0.9 M_\odot$ and (c) $1.2 M_\odot$. The naive model does not account for cases in which the starshade leaves the stellar disc, in contrast to the adjusted model, which incorporate these cases via equation \ref{eqn:integrals}. Note that the adjustment is more significant for smaller stars.}
    \label{fig:shade_size}
\end{figure}

The size of the starshade grows with time as the stellar luminosity and radius increases as shown in Fig.~\ref{fig:shade_size}. If the civilization has tried to maintain a certain temperature for a long time, the shade will eventually be of similar size to the planet and therefore likely be observable. In Fig.~\ref{fig:shade_size} we have chosen stars of mass 0.7, 1.0 and 1.2$\mathrm{M}_\odot$ to visualise our model for the starshade size versus the naive model that ignores cases when the starshade leaves the stellar disc (Equation~\ref{eqn:naivearea}). Our results show that the difference between the adjusted and naive starshade size decreases with stellar mass. So, for the more massive stars, the starshade does not leave the stellar disc (as seen from the planetary limb) which means that the naive calculation is enough to accurately determine the starshade size.

\begin{figure}
\centering
	\includegraphics[scale=0.55]{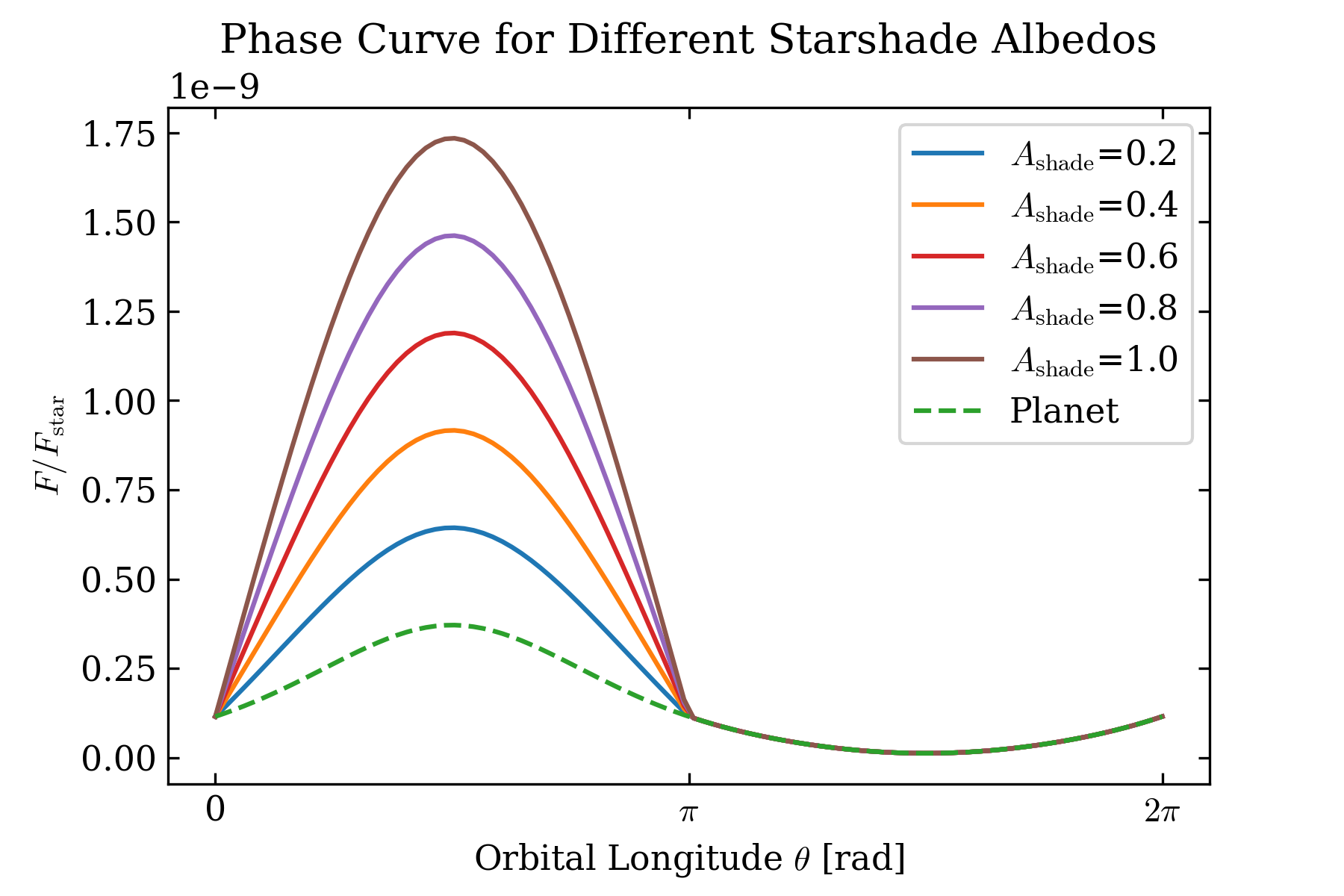}
    \caption{The total phase function (planet + starshade) for different starshade albedos. The phase function for the planet is also vizualized for reference. This plot was made assuming a $1 M_\odot$, 8 Gyr star with a $1 M_\oplus$ planet. The system was placed at a distance 8 pc away with an inclination of $i=\pi/3$. The starshade was assumed to be built at a target age of 3 Gyr, resulting in a starshade size of 0.906 $R_\oplus$ at current age.}
    \label{fig:phase_curve_albedo}
\end{figure}

The phase curve for an example starshade--planet system is shown in Fig.~\ref{fig:phase_curve_albedo} for different shade albedos, together with a typical planet phase curve without a starshade. Note that the starshade causes the amplitude of the phase curve to increase from $\sim 3 \times 10^{-10}$ to $>1 \times 10^{-9}$. The technosignature we expect is strongly dependent on how reflective the starshade is. Also prominent is the distinct shape of the starshade--planet curve compared to that of a typical planet. For half of the orbit, the starshade greatly amplifies the phase curve, while it does not contribute for the other half (when its unilluminated side is oriented towards the viewer). Similarly, the detectability of the technosignature will be dependent on the inclination -- the more face-on the system is, the smaller the starshade's projected area appears to the observer, and the weaker the amplification of the phase curve.

\begin{figure}
\centering
	\includegraphics[scale=0.55]{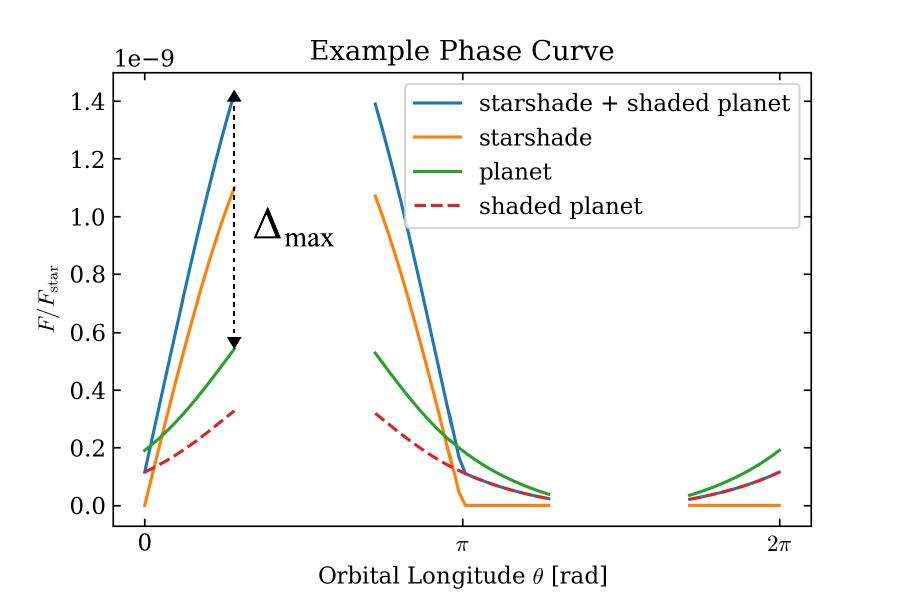}
    \caption{Example of a phase curve simulation for a $1 M_\odot$ star with a $1 M_\oplus$ planet. The system was placed at a distance 10 pc away with an inclination of $i=\pi/2$ (edge on), resulting in some parts of the system being covered by the IWA (which was assumed to be 60 mas). The starshade was introduced at a target age of 3 Gyr, resulting in a starshade size of 0.906 $R_\oplus$ at current age. $\Delta_\mathrm{max}$ is noted with a black dashed line, representing the maximum difference between the total planet + starshade phase curve and that of the planet without a starshade.}
    \label{fig:phase_curve_ex}
\end{figure}

Another example phase curve is shown in Fig.~\ref{fig:phase_curve_ex}, where we have now included the curve for each component in the system individually. The total phase curve from the starshade--planet system is also shown, and here we account for a reduction of the light reflected by the planet because of the starshade, as well as remove parts of the phase curve inside the IWA. To quantify the difference between the phase curves, we define $\Delta_\mathrm{max}$, which is the maximum observed difference (i.e. just outside the IWA) between the total phase curve and that of a typical planet. 

These specific cases show that the starshade may need to be large, comparable in size with the planet, and, if it has a high albedo, it will dominate the phase curve around opposition. Amplitudes are expected to be much larger than the HWO $1\sigma$ precision. Next, we explore the detectability of starshades among the HWO targets in more detail.

\subsection{Full population of HWO targets}

\label{sec:full_pop}

Moving on to the full population of HWO targets, we applied the starshade simulation to all target stars in the list provided by \cite{Harada2024}. 

For each star that meets the conditions defined in Section~\ref{sec:Method}, we calculate $\Delta_\mathrm{max}$ for 100 randomly drawn inclinations per age in the relevant subset among the target ages $t=[1,3,5,7,9]$ Gyr, assuming a starshade geometric albedo of 0.9. In principle, we would like to fit models of a planet alone versus a planet plus a starshade, to see whether the latter can be statistically distinguished from the former. However, this would rely on assumptions about the HWO observing strategy, such as number of visits and cadence, which is not yet determined. Therefore, in order to quantify the results we have chosen to look at the maximum phase curve difference $\Delta_\mathrm{max}$ between the starshade + shaded planet and the planet alone. The stars can then be ranked from highest to lowest $\left<\Delta_\mathrm{max}\right>$, where the $\left<\Delta_\mathrm{max}\right>$ we show is the mean over all target ages and inclination angles (up to 500 simulations per star). The average $\Delta_\mathrm{max}$ will naturally depend on the stellar luminosity, stellar age and how much of the system is covered by the IWA. As such, a higher value on $\left<\Delta_\mathrm{max}\right>$ means a more distinct technosignature in the phase curve simulations for that star. 

\begin{figure}
    \begin{subfigure}{.49\textwidth}
        \centering
        \includegraphics[scale=0.55]{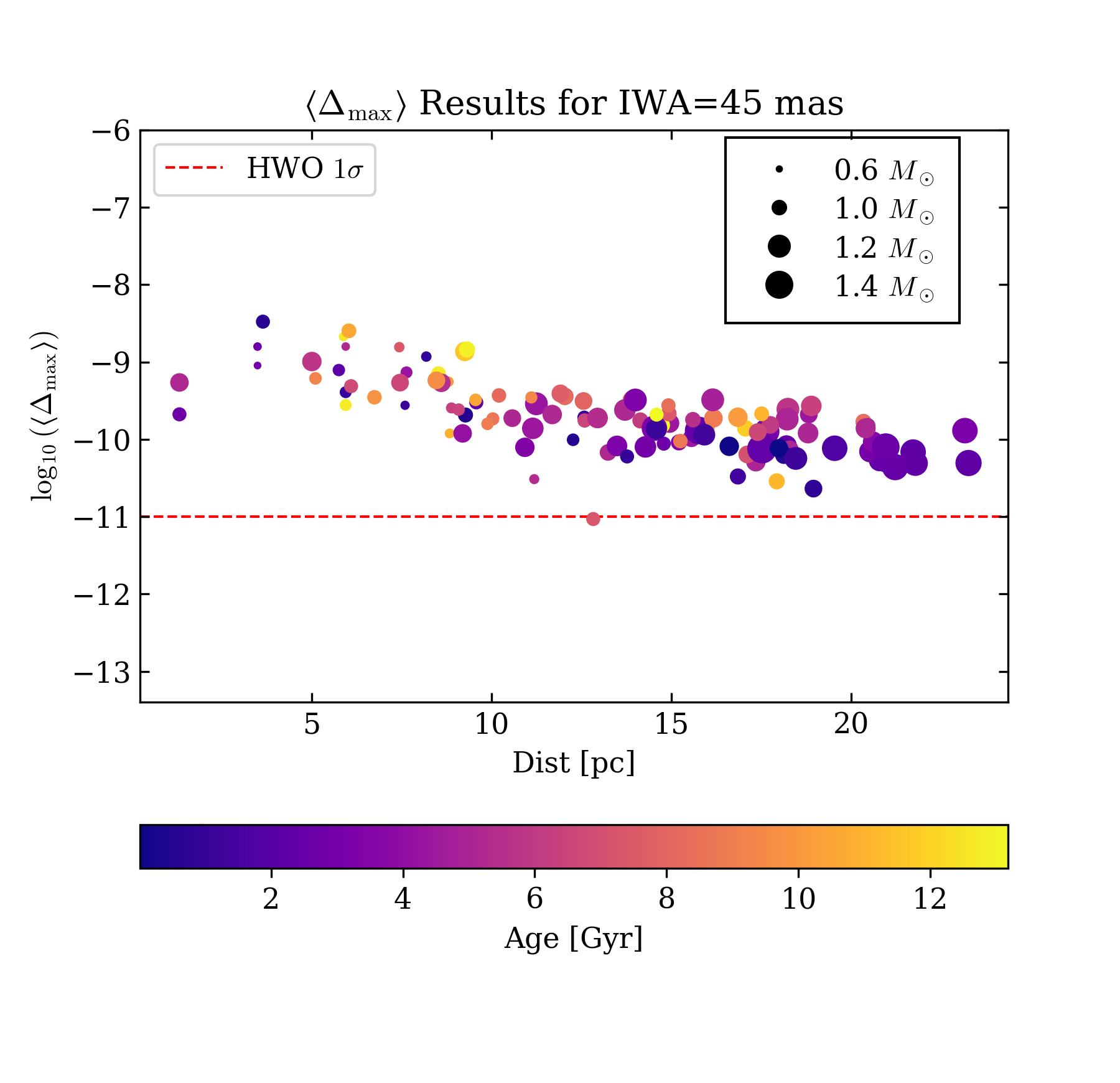}
        \vspace{-1cm}
        \caption{}
    \end{subfigure}\hfill
    \begin{subfigure}{.49\textwidth}
        \centering
        \includegraphics[scale=0.55]{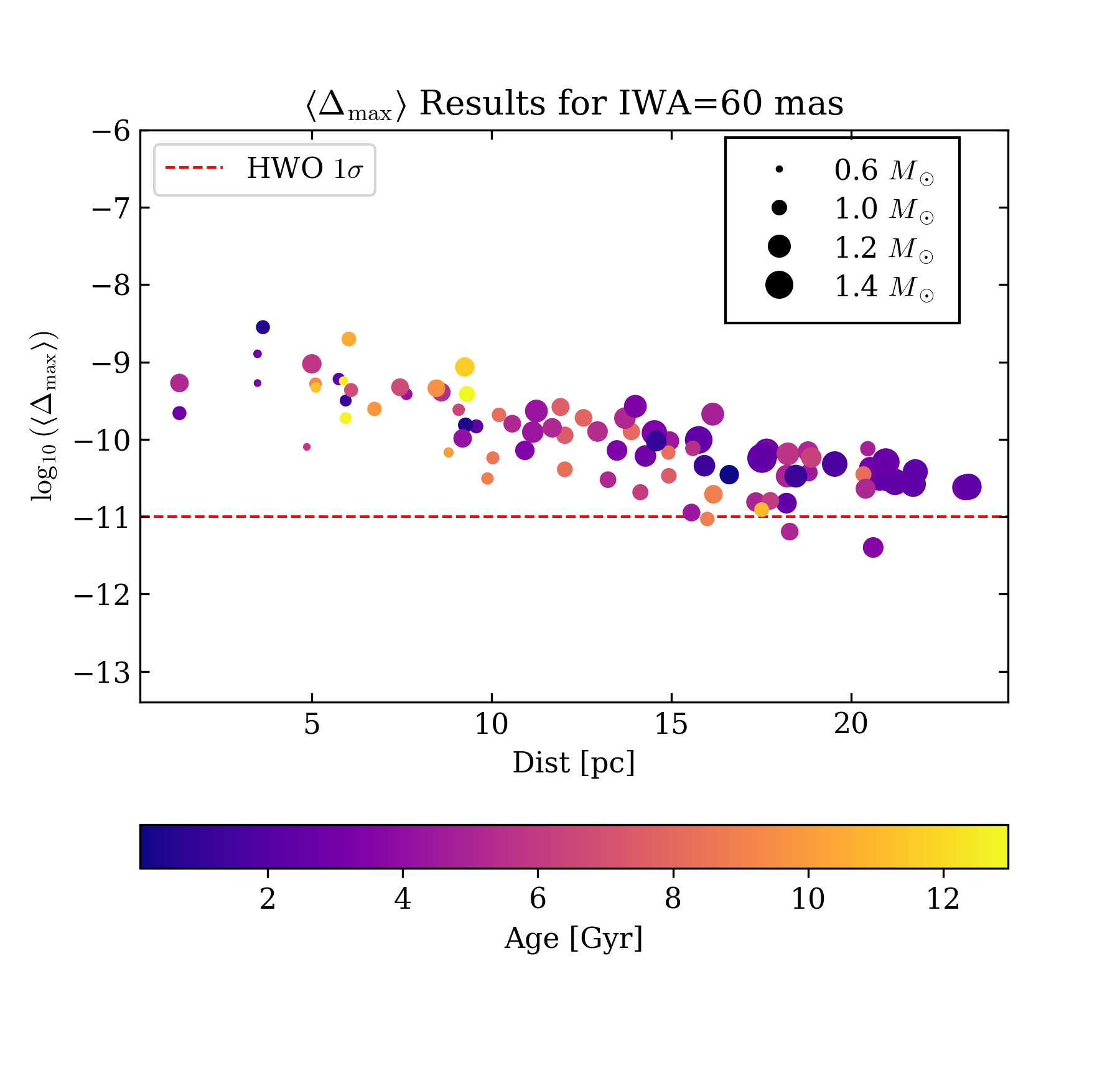}
        \vspace{-1cm}
        \caption{}
    \end{subfigure}
    \caption{$\left<\Delta_\mathrm{max}\right>$ ($\Delta_\mathrm{max}$ averaged over all inclinations and the relevant subset of target ages for each star) as a function of distance for (a) IWA=45 mas and (b) IWA=60 mas. $\left<\Delta_\mathrm{max}\right>$ decreases with increasing distance from the target star.}
    \label{fig:HWO_dist}
\end{figure}

\begin{figure}
    \begin{subfigure}{.49\textwidth}
        \centering
        \includegraphics[scale=0.55]{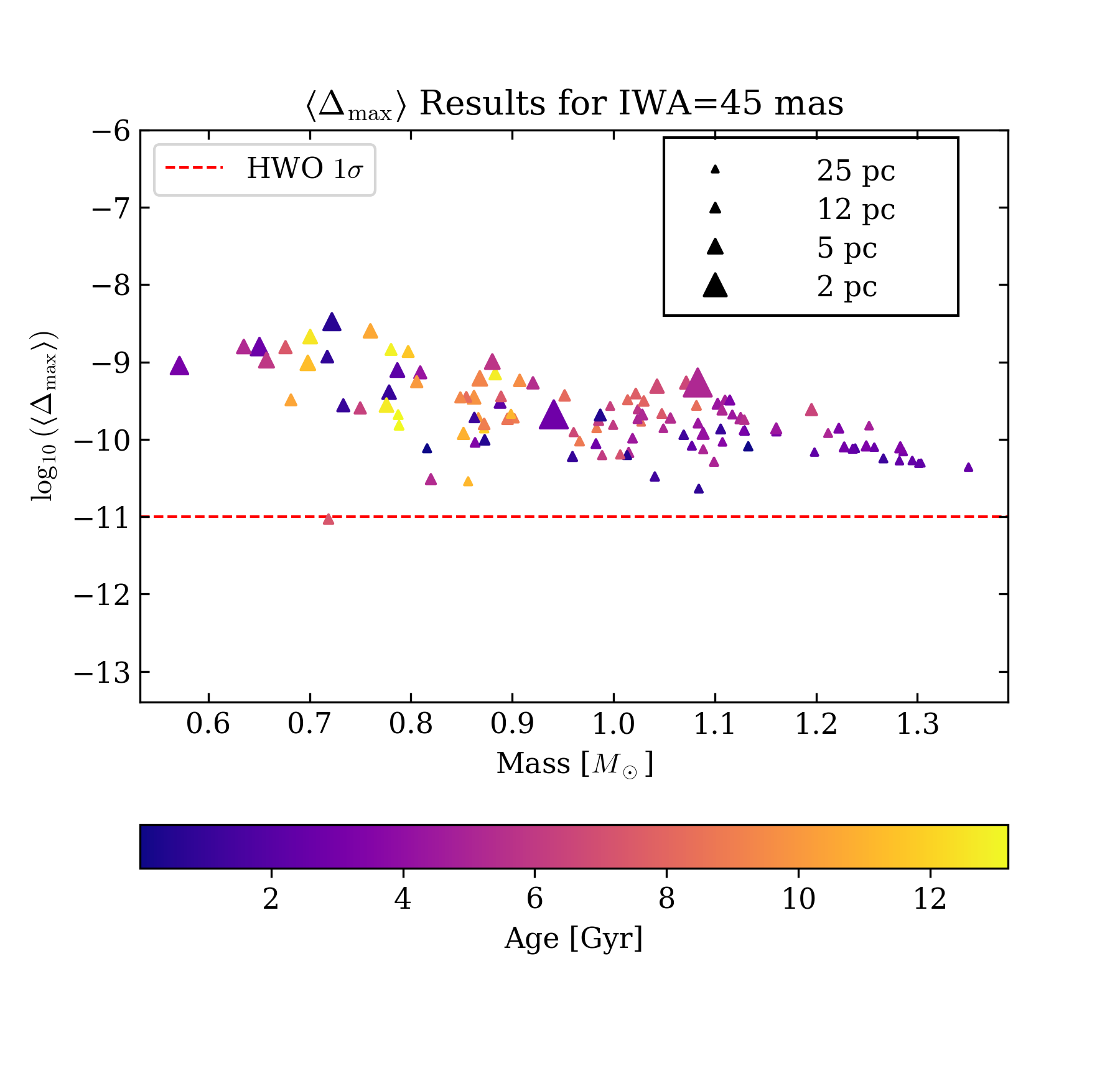}
        \vspace{-1cm}
        \caption{}
    \end{subfigure}\hfill
    \begin{subfigure}{.49\textwidth}
        \centering
        \includegraphics[scale=0.55]{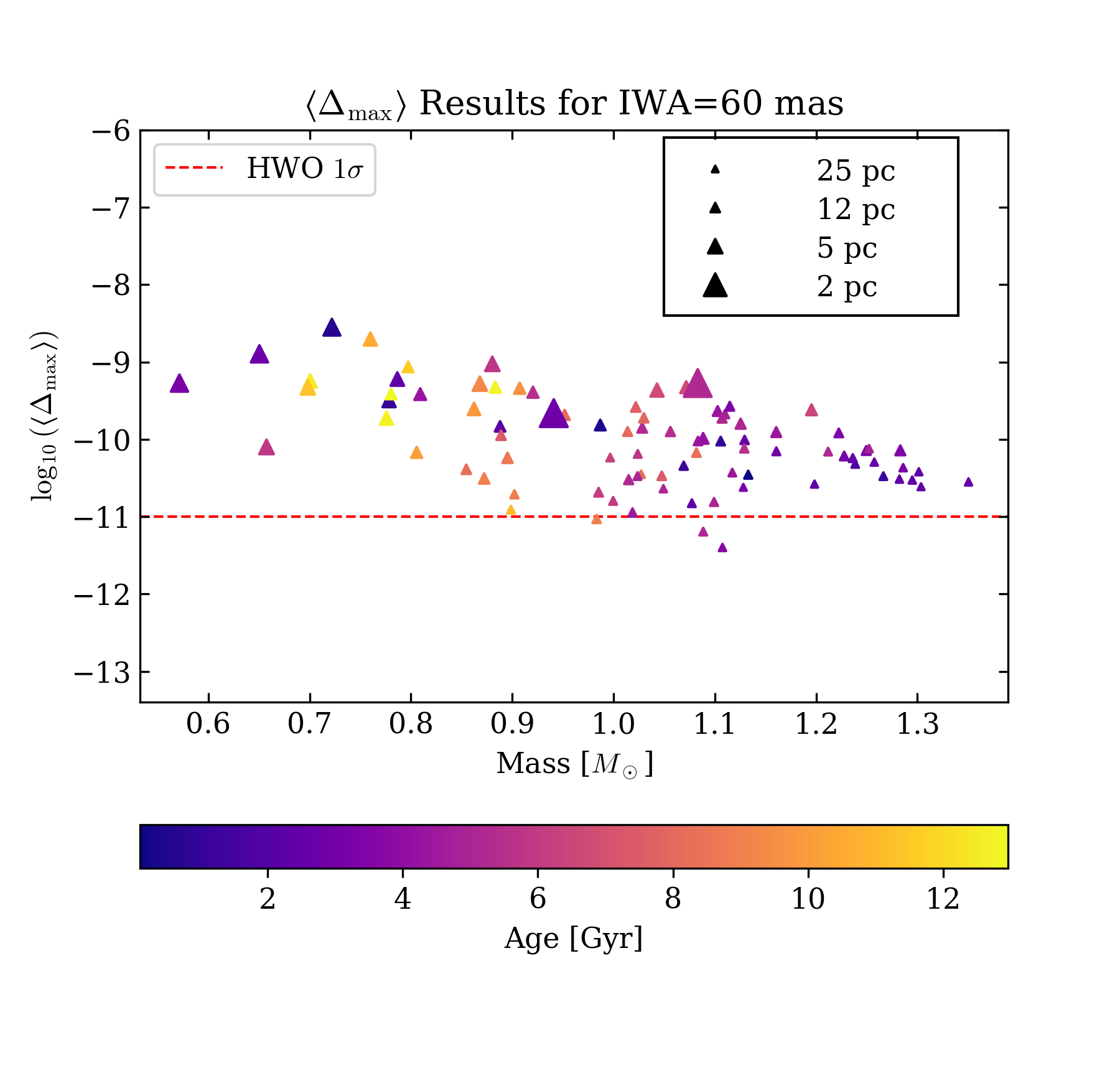}
        \vspace{-1cm}
        \caption{}
    \end{subfigure}
    \caption{$\left<\Delta_\mathrm{max}\right>$ ($\Delta_\mathrm{max}$ averaged over all inclinations and the relevant subset of target ages for each star) as a function of stellar mass for (a) IWA=45 mas and (b) IWA=60 mas. The lower-mass stars generally exhibit larger values of $\left<\Delta_\mathrm{max}\right>$.}
    \label{fig:HWO_mass}
\end{figure}

\begin{figure}
    \begin{subfigure}{.49\textwidth}
        \centering
        \includegraphics[scale=0.55]{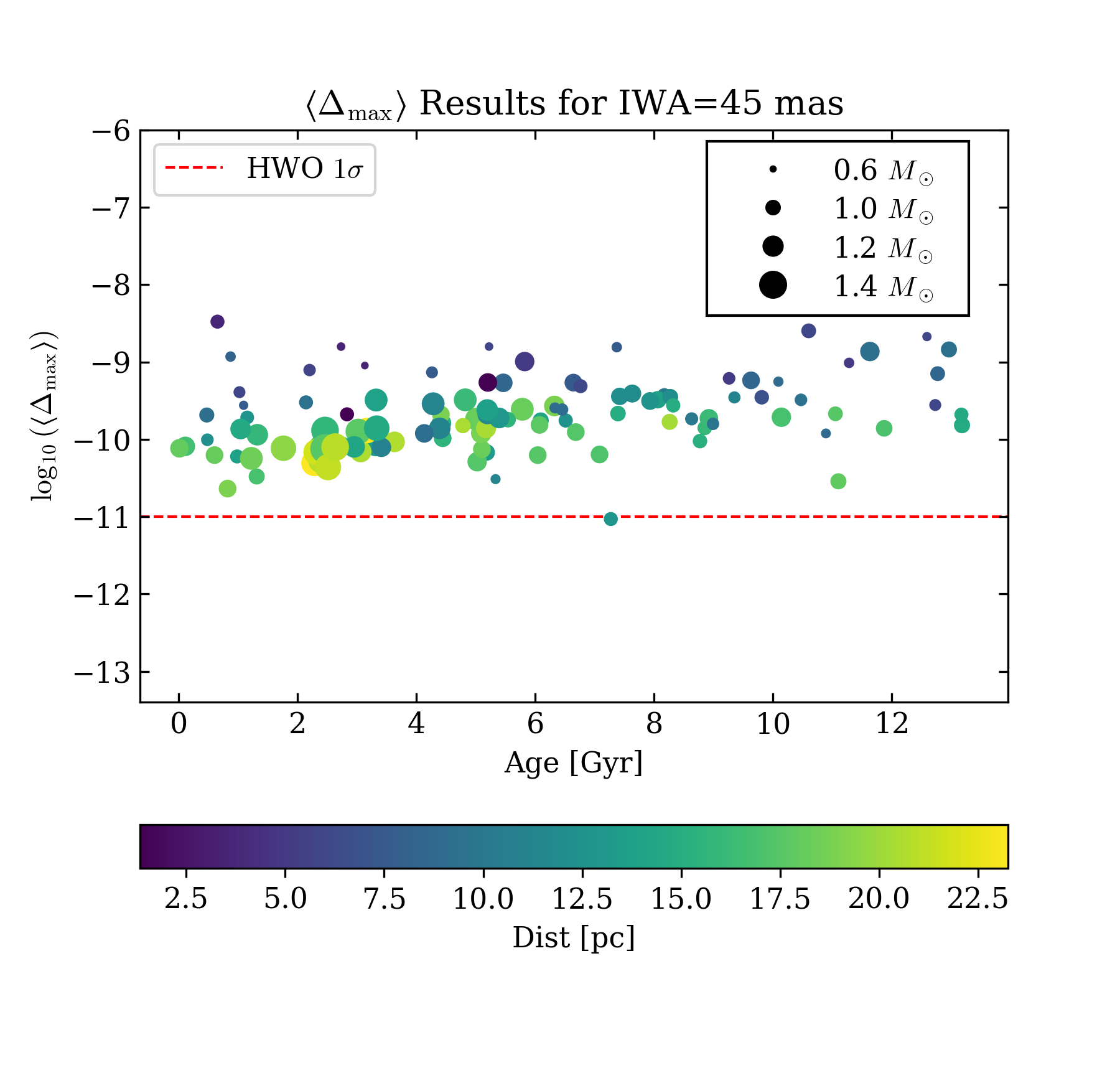}
        \vspace{-1cm}
        \caption{}
    \end{subfigure}\hfill
    \begin{subfigure}{.49\textwidth}
        \centering
        \includegraphics[scale=0.55]{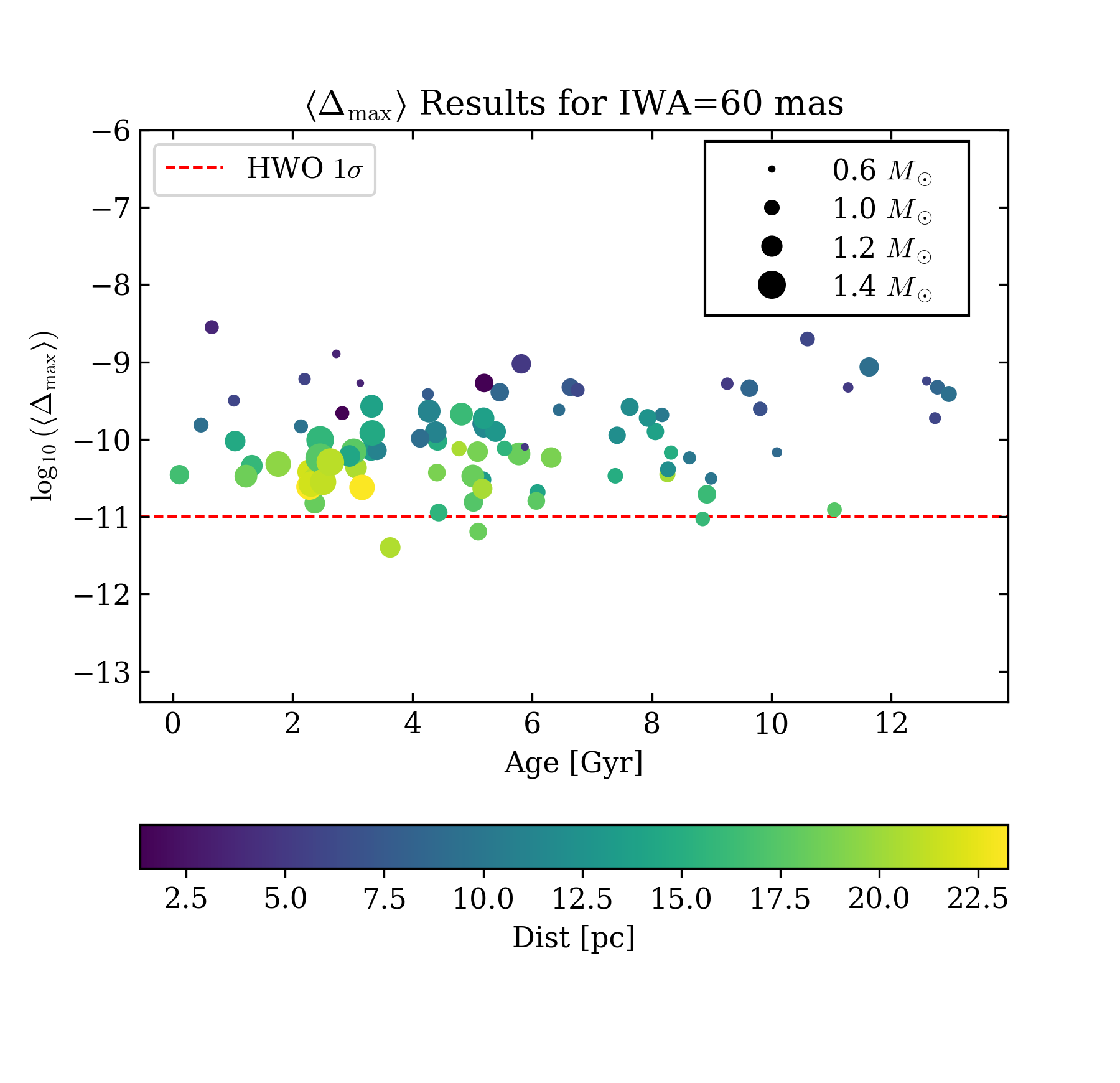}
        \vspace{-1cm}
        \caption{}
    \end{subfigure}
    \caption{$\left<\Delta_\mathrm{max}\right>$ ($\Delta_\mathrm{max}$ averaged over all inclinations and the relevant subset of target ages for each star) as a function of stellar age for (a) IWA=45 mas and (b) IWA=60 mas. The older stars tend to have large values of $\left<\Delta_\mathrm{max}\right>$ while the younger stars tend towards lower $\left<\Delta_\mathrm{max}\right>$-values.}
    \label{fig:HWO_age}
\end{figure}

The results from the simulations encompassing all HWO target stars are shown in Fig.~\ref{fig:HWO_dist},~\ref{fig:HWO_mass} and~\ref{fig:HWO_age}. Each of the figures consists of two panels: one for an IWA of 45 mas and one for an IWA of 60 mas. The $\left<\Delta_\mathrm{max}\right>$ values are also listed for each star in Table \ref{tab:HWO_stars} in Appendix A, together with the derived stellar masses, ages and other relevant parameters for the simulations. 

While it might seem unjustified to draw any conclusion about which targets are more likely to have potential technosignatures, it should be noted that a higher $\left<\Delta_\mathrm{max}\right>$ favours older (see Fig.~\ref{fig:HWO_age}), lower-mass (see Fig.~\ref{fig:HWO_mass}) stars. We can explain this because planets around young stars might not have had enough time to develop intelligent life and high-mass stars live much shorter lives, which limits the time-frame for intelligent life to develop. Nevertheless, the strongest dependency for $\left<\Delta_\mathrm{max}\right>$ is on the distance $D_*$ from Earth to the target star as seen in Fig.~\ref{fig:HWO_dist}. For nearby stars, less of the planet's orbit is covered by the IWA which means that the technosignature will be easier to detect. Additionally, the HWO sample has the majority of the low-mass stars at short distances, which means we get an additional bias in the data. 

Our results also highlight the significance of a small IWA for better detecting planets with small angular separation from their host star. We compare our calculated $\left<\Delta_\mathrm{max}\right>$ with the single-epoch precision foreseen for HWO of $1\sigma=10^{-11}$. With an IWA of 60 mas, we find a $\left<\Delta_\mathrm{max}\right> >1\sigma$ for ${70.8}\%$ of the HWO target stars above 1.5 Gyr. A narrower IWA of 45 mas increases this percentage to $96.7\%$.

\section{Discussion}
\label{sec:disc}

\subsection{Detectability of starshades}
\label{sec:detectability}

From the results, we conclude that a starshade can be detected, not only in transits as shown by \cite{Gaidos2017}, but also in direct imaging phase curves with near-future technology. While the transit method looks at a large number of targets with a small probability of detection, imaging looks at a small number of targets with a large probability of detection (should such structures exist). The phase curves have the advantage over transits that they do not require a specific transit geometry. Unless the orbit is very close to face-on, the phase curves should provide us with a distinguishable technosignature. 

The possibility of detecting a technosignature will depend on the IWA of the telescope. \cite{Vaughan2023} shows that an IWA smaller than the originally proposed 62 mas would benefit the detection of atmospheric and ocean features of the directly imaged exoplanets. Similarly, we have shown that a smaller IWA would also increase the possibility of detecting technosignatures; with an IWA of 60 mas, larger parts of the phase curves are not detectable, and, for some low-mass stars, the EED is completely inside the IWA. 

The technosignature is mainly expected to deviate from a `normal' phase curve by an anomalous shape in terms of the geometry, being more peaked towards opposition. The phase curve will also be greatly amplified since the concept presented here assumes a highly-reflective (high albedo) starshade. However, as the planet is not resolved, a phase curve that is only amplified can be explained by several natural phenomena. For instance, a larger planet gives a larger amount of reflected light. The same holds for a planet with ring structures or a planet with a large amount of reflective clouds in the atmosphere. Moreover, it is important to acknowledge that a civilisation may choose a low-albedo material for stealth, if they do not want it to be detected.

It is necessary to evaluate whether clouds, rings or moons of a planet could give a similar expression in terms of phase curve geometry to that of a starshade. The phase curves of planets with ring structures or orbiting moons are not expected to be similar to our results \citep[see e.g.][for comparison]{Dyudina2005, Luger2022}. However, clouds present a greater challenge.  Comparing our results to simulated phase curves of reflective clouds as in \cite{Hamill2024}, \cite{Mahpathra2023} and \cite{Mayorga2019}, we see that generally those do not exhibit a similar shape to that of the starshade phase curve. However, some models in \cite{Mayorga2019}, specifically for small-particle, tall, lofted clouds (i.e. clouds that are vertically thick) can give steep phase curves with abrupt changes in flux ratio, similarly to what is seen for the starshade in Fig. \ref{fig:phase_curve_ex}, and therefore such models should be included when evaluating technosignature claims in reflected light direct imaging attempts. We note, however, that the planets in \cite{Mayorga2019} are all too close to their star to be in the habitable zone and are all larger than the Earth. It is therefore unsure if such phenomena would occur on terrestrial planets in the habitable zone. \cite{Mahpathra2023} shows phase curve examples of exo-Earths and these are noticeably different compared to our starshade + exo-earth examples. Another potential cause of a false positive we can propose is a planet with an extreme axial tilt that has a polar ice-cap facing towards the observer at opposition. This would in principle give a similar geometry to that presented here, as the reflective patch is only visible in parts of the orbit and remains in the same location on the planet as seen from an observer. Another alternative could be a patch of very reflective clouds located at the substellar point, which faces away from the observer towards conjunction.

While we are considering starshades at $\mathcal{L}_1$ that are not resolvable from the planet, if they are made thin enough to offset them significantly as in \cite{Gaidos2017}, they may appear visually resolved. Such a co-rotating object would be clearly artificial as the L1 point (or equivalent, correcting for radiation pressure) is not stable. This is an example of technosignatures being unambiguous, as argued by \cite{Wright2022}.

\subsection{Impact of assumptions and limitations}
\label{sec:impactofassumptions}

During this study, we have made a number of simplifying assumptions and focused on a general estimation of observable effects of starshades in reflected light. For instance, we treat the star as a blackbody disc and limb darkening is ignored. If these aspects were taken into consideration, we would get a more precise result of what an actual phase curve would look like; however, we did not consider these things necessary to demonstrate the distinct shape the technosignature would exhibit in the phase curve. 

In the target list by \cite{Harada2024} we noted that the mass of the stars do not always agree with their current luminosity, age and stellar type and therefore chose to estimate the masses (and other necessary parameters) simply but consistently from the effective temperature, luminosity and metallicity provided in the list. Nevertheless, by doing so we note that some of our stars have been assigned an age greater than the age of the universe. Since the HWO target list contains measured parameters from different sources for each star it is difficult to comment on this since the source of error can be different depending on which star we are looking at. However, our age estimation is solely from isochrone fitting, which could lead to inaccurate results, particularly for stars thought to be young on other grounds, such as rapid rotation, high lithium abundance, or magnetic activity.

With our new stellar ages, 80 of the stars are now younger and 83 older compared to those found by SED-fitting in \cite{Harada2024}. Moreover, half of the stars have ages that are significantly different from those in \cite{Harada2024} -- having either been increased or decreased by more than $50 \%$. We do not claim that these new ages are closer to the truth compared to those in the utilised target list, but they had to be adjusted to ensure consistency with other parameters during our evolution simulations.

As the planet moves behind the star, a flux dip is expected in the phase curve. It was not included in the simulations because that part of the curve is covered by the inner working angle of the telescope. Similarly, as the planet moves in front of the star \citep[the transit case discussed by][]{Gaidos2017}, it will not be seen for a close to edge-on orbit since it lies inside the IWA. Also worth mentioning is that, at some point (for a close to edge-on orbit), the starshade and planet align with each other with respect to the observer. This would mean that only the starshade is seen and hence the planet should be subtracted from the phase curve in those cases. However, this is still expected to occur within the IWA and is therefore not of importance to our results. 

Since $\Delta_\mathrm{max}$ occurs at the border of the IWA, this work assumes that the single shot precision of $1\sigma$ is achievable in this region. It should be noted that, in reality, the $1\sigma$ precision gets worse as one moves towards the star because, close to the IWA, there is usually still some extra scattered light present. This noise could mean that the $1\sigma$ precision we assume is simply not reachable just at the border of the IWA.

Another point to consider is that technology can be present on planets that are not necessarily habitable \citep[see, e.g.,][]{Wright2022}. Humanity started its technological advances on Earth, but these efforts have now expanded to other planets in the solar system. In addition, technology can outlive its creators. It is therefore important to note that technosignatures such as starshades may be abundant not only in direct relation to the planet that originally hosted the technologically advanced civilization. The time needed for life to evolve is, however, difficult to evaluate since we only have one data point for this -- life on planet Earth \citep[see e.g.][]{Mckay2014}. Also, this data point is subject to strong anthropic bias. The same holds regarding the nature of potential technosignatures -- our technosignature searches are based on existing human technology \citep{Wright2022}, or theoretical concepts we have imagined but not yet built. It is troublesome to draw any conclusion about what technology would look like for other species. Not considered in this study are potential costs, the method of construction, and the potential distortion to the planet's climate that a starshade would possibly impose. A discussion on the public perceptions of the benefits and risks of solar geoengineering can, for instance, be found in \cite{Low2024}. 

\subsection{Alternative starshade designs}
\label{sec:designs}

During this study, we have limited the geometry of the starshade to the simple, circular case. We have shown that this design may not be ideal because it gives a different amount of flux received on different parts of the planet, which could lead to serious climate issues. For the case of Earth, we would receive more flux at the North Pole and the South Pole if we were to use the circular starshade. Since the Earth spins around its own axis, the higher flux at the eastern/western parts of the planetary disc
is not as significant averaged over a diurnal rotation period. The poles would, in contrast, constantly receive more flux and this could, for instance, result in extensive melting of the polar ice. We, consequently have a few suggestions of other geometries that may be more suitable:
\begin{itemize}
    \item A rectangular starshade extending above/below the solar disc by a suitable amount would reduce the shading effect on different parts of the planet. 
    \item A "net", of filling factor $<1$, larger than the solar disc as seen from the planet, would block out a roughly equal amount of light accross the planet.
    \item Several "small" starshades in orbit of the planet. This alternative may be more within reach of today's technology. Reflectors in an Earth–orbit have been introduced in a few other studies \citep[e.g.][]{Govindsamy2000, Salazar2016}, albeit with the purpose of preventing anthropogenic climate change. This would likely lead to a very different phase curve to that presented in this paper. 
\end{itemize}
Ultimately, the phase curve signature will be strongly dependent on the design of the starshade itself. For instance, it will differ depending on whether the shade is reflecting light straight away from the planet or if it is letting it through at some scattering angle. Hence, it is difficult to generalize what phase curves signatures would look like for artificial structures such as starshades. However, we have shown that even a simple Lambertian reflector has a noticeable effect on the phase curve. 

\subsection{Establishing a technosignature framework}
\label{sec:framework}
Detection of a technosignature would be of tremendous cultural and philosophical import. Therefore, any claimed detection will need to be handled carefully. 
\cite{Lingam2023} showcases a Bayesian approach for assesing technosignatures, ensuring they are not only resistant to false positives but also have a reasonable prior probability of existance. \cite{Meadows2022} outline a standardised framework for biosignature detection, arguing that multiple lines of evidence and a rigorous assessment process are necessary to make credible claims of extraterrestrial life. A similar approach should be adopted in technosignature research to enhance the robustness of detection and interpretation. It may be beneficial to structure technosignature evaluation in a similar way to that presented by \cite{Meadows2022}: Level 1, focusing on the detection and identification of a potential technosignature before committing significant resources, followed by Level 2, which would require broader community efforts and interdisciplinary missions to assess whether an abiotic or non-technological origin is the more likely explanation.

If a detection of similar nature to that presented in this paper was to occur, it would be crucial to assess how it aligns with these two levels. It is important to notice, as pointed out by \cite{Wright2022}, that technosignatures can be unambiguous, unlike lots of biosignatures. On the other hand, while biosignatures are direct indicators of life, technosignatures are indirect; technology has the possibility of outliving its creators (as discussed in Section~\ref{sec:impactofassumptions}) and can also be present at locations life does not inhabit; humanity is, for instance, sending rovers to Mars.

For Level 1 in the \cite{Meadows2022} framework, the results in our paper indicate that the technosignature would most likely lie far above the proposed contrast ratio precision of the HWO telescope. Even if we have a 1$\sigma$ detection per epoch, many observations would still be able to give us a robust result. The most important issues are likely to be ensuring a stable calibration (for instance regarding where the telescope is in its orbit) and confusion with background sources. 

When it is confirmed that an authentic signal has been detected, it is necessary to rule out any potential false positives as part of Level 2. As discussed in Section~\ref{sec:detectability}, phase curves of ring structures and moons are not generally of similar shape. Cloud structures must, however, be evaluated more carefully, for example with multi-band photometry. This since some cloud phenomena can cause unusually shaped, or extremely peaked phase curves \citep[see][and the discussion in Section \ref{sec:detectability}]{Mayorga2019}. As part of Level 2, one will also need to evaluate whether life is expected to evolve in the environment that produced the signal. Here, factors such as the age of the host star and the stability of the planetary system need to be taken into consideration. Once again, however, the anthropic bias that comes with only having one data point for life makes this question difficult to evaluate. But one could argue that, if the system seems to be stable and fairly old, with the planet inside the HZ, it would be good conditions for technologically advanced life to develop. Then, further information such as spectral measurements, multi-band photometry and thermal IR measurements of the planet could strengthen or diminish the claim, depending on e.g. elemental abundances of the planet. This question also bridges technosignatures and biosignatures, as biosignatures from the same system would help to validate the findings.  

Lastly, if humanity in the future considers building a starshade or a similar structure, either because of anthropogenic climate change or because of the increasing luminosity of the sun, it is important to have in mind that such structure may be detectable by potential extraterrestrial civilizations. Studies investigating at what distances the technosignatures of the Earth are detectable \citep[see e.g.][]{Sheikh2025} stands as crucial means for advancing this discussion. 

\section{Conclusions}
\label{sec:conclusions}
This study aimed at investigating how the technosignature of a starshade would appear in direct imaging observations in reflected light, where the starshade is constructed to offset the increasing MS luminosity of the star and maintain a habitable temperature on an inhabited planet. Using isochrone fitting and stellar evolution models, we simulated phase curve signatures of starshade--planet systems orbiting the proposed targets of the upcoming HWO mission. The prominence of the technosignatures was quantized by $\Delta_\mathrm{max}$, which represents the difference in the observed phase curve amplitudes between a starshade--planet system and a planet without a starshade.

To offset increasing MS luminosity over several Gyr, starshades may need to be comparable in radius to the planet. We find that HWO will be easily capable of detecting these large starshades. With an IWA of 60 mas, we find a $\left<\Delta_\mathrm{max}\right>$ greater than the $1\sigma$ single-epoch precision for $70.8\%$ of the HWO target stars older than 1.5 Gyr. A narrower IWA of 45 mas increases this percentage to $96.7\%$, indicating the significance of a sufficiently low IWA for detecting anomalies in exoplanet light curves. However, without further information on the size and albedo of the host planet, an amplified phase function is not enough to identify a technosignature. Instead, the distinct shape of the phase curve, which is peaked towards opposition, can serve as a key indicator of a technologically advanced civilization. 

Although this study is hypothetical in nature, examining potential technosignatures remains essential for establishing a robust framework for assessing any detection claims. Future efforts should focus on examining how different technosignatures as well as natural phenomena are expected to appear in phase curves to mitigate false positives.

\section*{Acknowledgements}
CIS acknowledges Lund University and Stockholm University for providing access to essential research materials and resources throughout the scope of this work.

The authors wish to thank the anonymous referee for constructive comments that helped to improve the paper.

AJM acknowledges support from the Swedish Research Council (Project Grant 2022-04043) and the Swedish National Space Agency (Career Grants 120/19C and 2023-00146).

\textit{Software:} \textsc{ezmist}, \footnote{Morgan Fouesneau \& contributors: \url{https://github.com/mfouesneau/ezmist}}, \textsc{numpy} \citep{Numpy}, \textsc{astropy} \citep{Astropy1, Astropy2, Astropy3}, \textsc{scipy} \citep{Scipy}, \textsc{matplotlib} \citep{Matplotlib} and \textsc{pandas} \citep{Pandas}. The diagrams in this article were created using \textsc{inkscape} (\url{https://inkscape.org}).

CIS also acknowledges the use of ChatGPT for minor code optimization suggestions. Every incorporation of the suggested methods was cross-checked and adjusted by the author before implementation. 

\textit{Author contributions:} This work started as a BSc thesis\footnote{\url{https://lup.lub.lu.se/student-papers/search/publication/9124347}}, written by CIS and supervised by AJM. AJM conceived the original project. CIS has developed the code for the simulations and adjusted it according to AJM's advice. CIS has also written the majority of the manuscript, once again incorporating suggestions of improvement from AJM. 
\section*{Data Availability}
The software developed by the authors and the resulting data are available on Zenodo, at \url{https://doi.org/10.5281/zenodo.17175107} \citep{Skoglund_2025}.



\bibliographystyle{mnras}
\bibliography{example} 

\begin{thebibliography}{}
\makeatletter
\relax
\def\mn@urlcharsother{\let\do\@makeother \do\$\do\&\do\#\do\^\do\_\do\%\do\~}
\def\mn@doi{\begingroup\mn@urlcharsother \@ifnextchar [ {\mn@doi@} {\mn@doi@[]}}
\def\mn@doi@[#1]#2{\def\@tempa{#1}\ifx\@tempa\@empty \href {http://dx.doi.org/#2} {doi:#2}\else \href {http://dx.doi.org/#2} {#1}\fi \endgroup}
\def\mn@eprint#1#2{\mn@eprint@#1:#2::\@nil}
\def\mn@eprint@arXiv#1{\href {http://arxiv.org/abs/#1} {{\tt arXiv:#1}}}
\def\mn@eprint@dblp#1{\href {http://dblp.uni-trier.de/rec/bibtex/#1.xml} {dblp:#1}}
\def\mn@eprint@#1:#2:#3:#4\@nil{\def\@tempa {#1}\def\@tempb {#2}\def\@tempc {#3}\ifx \@tempc \@empty \let \@tempc \@tempb \let \@tempb \@tempa \fi \ifx \@tempb \@empty \def\@tempb {arXiv}\fi \@ifundefined {mn@eprint@\@tempb}{\@tempb:\@tempc}{\expandafter \expandafter \csname mn@eprint@\@tempb\endcsname \expandafter{\@tempc}}}

\bibitem[\protect\citeauthoryear{{Akeson} et~al.,}{{Akeson} et~al.}{2019}]{Akeson2019}
{Akeson} R.,  et~al., 2019, \mn@doi [arXiv e-prints] {10.48550/arXiv.1902.05569}, p. arXiv:1902.05569

\bibitem[\protect\citeauthoryear{{Amiri} \& {Loeb}}{{Amiri} \& {Loeb}}{2024}]{Amiri2024}
{Amiri} A.,  {Loeb} A.,  2024, in American Astronomical Society Meeting Abstracts. p. 159.04

\bibitem[\protect\citeauthoryear{{Angel}}{{Angel}}{2006}]{Angel2006}
{Angel} R.,  2006, \mn@doi [Proceedings of the National Academy of Science] {10.1073/pnas.0608163103}, 103, 17184

\bibitem[\protect\citeauthoryear{{Astropy Collaboration} et~al.,}{{Astropy Collaboration} et~al.}{2013}]{Astropy1}
{Astropy Collaboration} et~al., 2013, \mn@doi [\aap] {10.1051/0004-6361/201322068}, \href {http://adsabs.harvard.edu/abs/2013A%26A...558A..33A} {558, A33}

\bibitem[\protect\citeauthoryear{{Astropy Collaboration} et~al.,}{{Astropy Collaboration} et~al.}{2018}]{Astropy2}
{Astropy Collaboration} et~al., 2018, \mn@doi [\aj] {10.3847/1538-3881/aabc4f}, \href {https://ui.adsabs.harvard.edu/abs/2018AJ....156..123A} {156, 123}

\bibitem[\protect\citeauthoryear{{Astropy Collaboration} et~al.,}{{Astropy Collaboration} et~al.}{2022}]{Astropy3}
{Astropy Collaboration} et~al., 2022, \mn@doi [\apj] {10.3847/1538-4357/ac7c74}, 935, 167

\bibitem[\protect\citeauthoryear{{Baum}, {Low}  \& {Sovacool}}{{Baum} et~al.}{2022}]{Baum2022}
{Baum} C.~M.,  {Low} S.,   {Sovacool} B.~K.,  2022, \mn@doi [Renewable and Sustainable Energy Reviews] {10.1016/j.rser.2022.112179}, 158, 112179

\bibitem[\protect\citeauthoryear{{Bruna}, {Cowan}, {Sheffler}, {Haggard}, {Bourdon}  \& {M{\^a}lin}}{{Bruna} et~al.}{2023}]{Bruna2023}
{Bruna} M.,  {Cowan} N.~B.,  {Sheffler} J.,  {Haggard} H.~M.,  {Bourdon} A.,   {M{\^a}lin} M.,  2023, \mn@doi [\mnras] {10.1093/mnras/stac3521}, 519, 460

\bibitem[\protect\citeauthoryear{{Chauvin}, {Lagrange}, {Dumas}, {Zuckerman}, {Mouillet}, {Song}, {Beuzit}  \& {Lowrance}}{{Chauvin} et~al.}{2004}]{Chauvin2004}
{Chauvin} G.,  {Lagrange} A.~M.,  {Dumas} C.,  {Zuckerman} B.,  {Mouillet} D.,  {Song} I.,  {Beuzit} J.~L.,   {Lowrance} P.,  2004, \mn@doi [\aap] {10.1051/0004-6361:200400056}, 425, L29

\bibitem[\protect\citeauthoryear{{Choi}, {Dotter}, {Conroy}, {Cantiello}, {Paxton}  \& {Johnson}}{{Choi} et~al.}{2016}]{Choi2016}
{Choi} J.,  {Dotter} A.,  {Conroy} C.,  {Cantiello} M.,  {Paxton} B.,   {Johnson} B.~D.,  2016, \mn@doi [\apj] {10.3847/0004-637X/823/2/102}, 823, 102

\bibitem[\protect\citeauthoryear{{Dotter}}{{Dotter}}{2016}]{Dotter2016}
{Dotter} A.,  2016, \mn@doi [\apjs] {10.3847/0067-0049/222/1/8}, 222, 8

\bibitem[\protect\citeauthoryear{{Dubey}, {Kopparapu}, {Ercolano}  \& {Molaverdikhani}}{{Dubey} et~al.}{2025}]{Dubey2025}
{Dubey} D.,  {Kopparapu} R.,  {Ercolano} B.,   {Molaverdikhani} K.,  2025, \mn@doi [The Planetary Science Journal] {10.3847/PSJ/ad98eb}, 6, 4

\bibitem[\protect\citeauthoryear{{Dyudina}, {Sackett}, {Bayliss}, {Seager}, {Porco}, {Throop}  \& {Dones}}{{Dyudina} et~al.}{2005}]{Dyudina2005}
{Dyudina} U.~A.,  {Sackett} P.~D.,  {Bayliss} D. D.~R.,  {Seager} S.,  {Porco} C.~C.,  {Throop} H.~B.,   {Dones} L.,  2005, \mn@doi [\apj] {10.1086/426050}, 618, 973

\bibitem[\protect\citeauthoryear{{Fuglesang} \& {de Herreros Miciano}}{{Fuglesang} \& {de Herreros Miciano}}{2021}]{Fuglesang2021}
{Fuglesang} C.,  {de Herreros Miciano} M.~G.,  2021, \mn@doi [Acta Astronautica] {10.1016/j.actaastro.2021.04.035}, 186, 269

\bibitem[\protect\citeauthoryear{{Gaidos}}{{Gaidos}}{2017}]{Gaidos2017}
{Gaidos} E.,  2017, \mn@doi [\mnras] {10.1093/mnras/stx1078}, 469, 4455

\bibitem[\protect\citeauthoryear{{Gaudi} et~al.,}{{Gaudi} et~al.}{2020}]{HABEX2020}
{Gaudi} B.~S.,  et~al., 2020, \mn@doi [arXiv e-prints] {10.48550/arXiv.2001.06683}, p. arXiv:2001.06683

\bibitem[\protect\citeauthoryear{{Gough}}{{Gough}}{1981}]{Gough1981}
{Gough} D.~O.,  1981, \mn@doi [\solphys] {10.1007/BF00151270}, 74, 21

\bibitem[\protect\citeauthoryear{{Govindasamy} \& {Caldeira}}{{Govindasamy} \& {Caldeira}}{2000}]{Govindsamy2000}
{Govindasamy} B.,  {Caldeira} K.,  2000, \mn@doi [\grl] {10.1029/1999GL006086}, 27, 2141

\bibitem[\protect\citeauthoryear{{Hamill}, {Johnson}, {Batalha}, {Nag}, {Gao}, {Adams}  \& {Kataria}}{{Hamill} et~al.}{2024}]{Hamill2024}
{Hamill} C.~D.,  {Johnson} A.~V.,  {Batalha} N.,  {Nag} R.,  {Gao} P.,  {Adams} D.,   {Kataria} T.,  2024, \mn@doi [\apj] {10.3847/1538-4357/ad7de6}, 976, 181

\bibitem[\protect\citeauthoryear{{Harada}, {Dressing}, {Kane}  \& {Ardestani}}{{Harada} et~al.}{2024}]{Harada2024}
{Harada} C.~K.,  {Dressing} C.~D.,  {Kane} S.~R.,   {Ardestani} B.~A.,  2024, \mn@doi [\apjs] {10.3847/1538-4365/ad3e81}, 272, 30

\bibitem[\protect\citeauthoryear{Harris et~al.,}{Harris et~al.}{2020}]{Numpy}
Harris C.~R.,  et~al., 2020, \mn@doi [Nature] {10.1038/s41586-020-2649-2}, 585, 357

\bibitem[\protect\citeauthoryear{{Hart}}{{Hart}}{1978}]{Hart1978}
{Hart} M.~H.,  1978, \mn@doi [\icarus] {10.1016/0019-1035(78)90021-0}, 33, 23

\bibitem[\protect\citeauthoryear{{Heng}, {Morris}  \& {Kitzmann}}{{Heng} et~al.}{2021}]{Heng2021}
{Heng} K.,  {Morris} B.~M.,   {Kitzmann} D.,  2021, \mn@doi [Nature Astronomy] {10.1038/s41550-021-01444-7}, 5, 1001

\bibitem[\protect\citeauthoryear{{Huang}}{{Huang}}{1959}]{Huang1959}
{Huang} S.-S.,  1959, American Scientist, 47, 397

\bibitem[\protect\citeauthoryear{Hunter}{Hunter}{2007}]{Matplotlib}
Hunter J.~D.,  2007, \mn@doi [Computing in Science \& Engineering] {10.1109/MCSE.2007.55}, 9, 90

\bibitem[\protect\citeauthoryear{{Kane} \& {Gelino}}{{Kane} \& {Gelino}}{2010}]{Kane_2010}
{Kane} S.~R.,  {Gelino} D.~M.,  2010, \mn@doi [\apj] {10.1088/0004-637X/724/1/818}, 724, 818

\bibitem[\protect\citeauthoryear{{Kayali}, {Haliki}, {Bas}  \& {Nemiroff}}{{Kayali} et~al.}{2024}]{Kayali2024}
{Kayali} O.,  {Haliki} E.,  {Bas} K.,   {Nemiroff} R.~J.,  2024, \mn@doi [arXiv e-prints] {10.48550/arXiv.2412.17086}, p. arXiv:2412.17086

\bibitem[\protect\citeauthoryear{{Kopparapu}, {Ramirez}, {SchottelKotte}, {Kasting}, {Domagal-Goldman}  \& {Eymet}}{{Kopparapu} et~al.}{2014}]{Kopparapu2014}
{Kopparapu} R.~K.,  {Ramirez} R.~M.,  {SchottelKotte} J.,  {Kasting} J.~F.,  {Domagal-Goldman} S.,   {Eymet} V.,  2014, \mn@doi [\apjl] {10.1088/2041-8205/787/2/L29}, 787, L29

\bibitem[\protect\citeauthoryear{{Kopparapu}, {Kofman}, {Haqq-Misra}, {Kopparapu}  \& {Lingam}}{{Kopparapu} et~al.}{2024}]{Kopparapu2024}
{Kopparapu} R.,  {Kofman} V.,  {Haqq-Misra} J.,  {Kopparapu} V.,   {Lingam} M.,  2024, \mn@doi [\apj] {10.3847/1538-4357/ad43d7}, 967, 119

\bibitem[\protect\citeauthoryear{{Lagrange} et~al.,}{{Lagrange} et~al.}{2009}]{Lagrange2009}
{Lagrange} A.~M.,  et~al., 2009, \mn@doi [\aap] {10.1051/0004-6361:200811325}, 493, L21

\bibitem[\protect\citeauthoryear{{Li}, {Hildebrandt}, {Kane}, {Zimmerman}, {Girard}, {Gonzalez-Quiles}  \& {Turnbull}}{{Li} et~al.}{2021}]{Li2021}
{Li} Z.,  {Hildebrandt} S.~R.,  {Kane} S.~R.,  {Zimmerman} N.~T.,  {Girard} J.~H.,  {Gonzalez-Quiles} J.,   {Turnbull} M.~C.,  2021, \mn@doi [\aj] {10.3847/1538-3881/abf831}, 162, 9

\bibitem[\protect\citeauthoryear{{Lingam}, {Haqq-Misra}, {Wright}, {Huston}, {Frank}  \& {Kopparapu}}{{Lingam} et~al.}{2023}]{Lingam2023}
{Lingam} M.,  {Haqq-Misra} J.,  {Wright} J.~T.,  {Huston} M.~J.,  {Frank} A.,   {Kopparapu} R.,  2023, \mn@doi [\apj] {10.3847/1538-4357/acaca0}, 943, 27

\bibitem[\protect\citeauthoryear{{Low}, {Fritz}, {Baum}  \& {Sovacool}}{{Low} et~al.}{2024}]{Low2024}
{Low} S.,  {Fritz} L.,  {Baum} C.~M.,   {Sovacool} B.~K.,  2024, \mn@doi [Communications Earth and Environment] {10.1038/s43247-024-01518-0}, 5, 352

\bibitem[\protect\citeauthoryear{{Luger}, {Agol}, {Bartoli{\'c}}  \& {Foreman-Mackey}}{{Luger} et~al.}{2022}]{Luger2022}
{Luger} R.,  {Agol} E.,  {Bartoli{\'c}} F.,   {Foreman-Mackey} D.,  2022, \mn@doi [\aj] {10.3847/1538-3881/ac4017}, \href {https://ui.adsabs.harvard.edu/abs/2022AJ....164....4L} {164, 4}

\bibitem[\protect\citeauthoryear{{Madhusudhan}}{{Madhusudhan}}{2019}]{Madhusudhan2019}
{Madhusudhan} N.,  2019, \mn@doi [\araa] {10.1146/annurev-astro-081817-051846}, 57, 617

\bibitem[\protect\citeauthoryear{{Madhusudhan} \& {Burrows}}{{Madhusudhan} \& {Burrows}}{2012}]{Madhusudhan2012}
{Madhusudhan} N.,  {Burrows} A.,  2012, \mn@doi [\apj] {10.1088/0004-637X/747/1/25}, 747, 25

\bibitem[\protect\citeauthoryear{{Mahapatra}, {Abiad}, {Rossi}  \& {Stam}}{{Mahapatra} et~al.}{2023}]{Mahpathra2023}
{Mahapatra} G.,  {Abiad} F.,  {Rossi} L.,   {Stam} D.~M.,  2023, \mn@doi [\aap] {10.1051/0004-6361/202243190}, 671, A165

\bibitem[\protect\citeauthoryear{{Mallama}, {Krobusek}  \& {Pavlov}}{{Mallama} et~al.}{2017}]{Mallama2017}
{Mallama} A.,  {Krobusek} B.,   {Pavlov} H.,  2017, \mn@doi [\icarus] {10.1016/j.icarus.2016.09.023}, 282, 19

\bibitem[\protect\citeauthoryear{{Mamajek} \& {Stapelfeldt}}{{Mamajek} \& {Stapelfeldt}}{2024}]{Mamajek2024}
{Mamajek} E.,  {Stapelfeldt} K.,  2024, \mn@doi [arXiv e-prints] {10.48550/arXiv.2402.12414}, \href {https://ui.adsabs.harvard.edu/abs/2024arXiv240212414M} {p. arXiv:2402.12414}

\bibitem[\protect\citeauthoryear{{Marois}, {Macintosh}, {Barman}, {Zuckerman}, {Song}, {Patience}, {Lafreni{\`e}re}  \& {Doyon}}{{Marois} et~al.}{2008}]{Marois2008}
{Marois} C.,  {Macintosh} B.,  {Barman} T.,  {Zuckerman} B.,  {Song} I.,  {Patience} J.,  {Lafreni{\`e}re} D.,   {Doyon} R.,  2008, \mn@doi [Science] {10.1126/science.1166585}, 322, 1348

\bibitem[\protect\citeauthoryear{{Mayorga}, {Batalha}, {Lewis}  \& {Marley}}{{Mayorga} et~al.}{2019}]{Mayorga2019}
{Mayorga} L.~C.,  {Batalha} N.~E.,  {Lewis} N.~K.,   {Marley} M.~S.,  2019, \mn@doi [\aj] {10.3847/1538-3881/ab29fa}, 158, 66

\bibitem[\protect\citeauthoryear{{McKay}}{{McKay}}{2014}]{Mckay2014}
{McKay} C.~P.,  2014, \mn@doi [Proceedings of the National Academy of Science] {10.1073/pnas.1304212111}, 111, 12628

\bibitem[\protect\citeauthoryear{{Meadows} et~al.,}{{Meadows} et~al.}{2022}]{Meadows2022}
{Meadows} V.,  et~al., 2022, \mn@doi [arXiv e-prints] {10.48550/arXiv.2210.14293}, p. arXiv:2210.14293

\bibitem[\protect\citeauthoryear{{Minunno}, {Andersson}  \& {Morrison}}{{Minunno} et~al.}{2023}]{Minunno2023}
{Minunno} R.,  {Andersson} N.,   {Morrison} G.~M.,  2023, \mn@doi [Earth Science Reviews] {10.1016/j.earscirev.2023.104431}, 241, 104431

\bibitem[\protect\citeauthoryear{Murray \& Dermott}{Murray \& Dermott}{2000}]{Murray2000}
Murray C.~D.,  Dermott S.~F.,  2000, Solar System Dynamics.
Cambridge University Press

\bibitem[\protect\citeauthoryear{{National Academies of Sciences, Medicine, and Engineering}}{{National Academies of Sciences, Medicine, and Engineering}}{2021}]{Decadal2020}
{National Academies of Sciences, Medicine, and Engineering} 2021, {Pathways to Discovery in Astronomy and Astrophysics for the 2020s}

\bibitem[\protect\citeauthoryear{{O'Malley-James}, {Greaves}, {Raven}  \& {Cockell}}{{O'Malley-James} et~al.}{2013}]{OMalley2013}
{O'Malley-James} J.~T.,  {Greaves} J.~S.,  {Raven} J.~A.,   {Cockell} C.~S.,  2013, \mn@doi [International Journal of Astrobiology] {10.1017/S147355041200047X}, 12, 99

\bibitem[\protect\citeauthoryear{{Parmentier} \& {Crossfield}}{{Parmentier} \& {Crossfield}}{2018}]{Parmentier2018}
{Parmentier} V.,  {Crossfield} I. J.~M.,  2018, in {Deeg} H.~J.,  {Belmonte} J.~A.,  eds, , Handbook of Exoplanets.
Springer, Cham., p.~116

\bibitem[\protect\citeauthoryear{{Paxton}, {Bildsten}, {Dotter}, {Herwig}, {Lesaffre}  \& {Timmes}}{{Paxton} et~al.}{2011}]{Paxton2011}
{Paxton} B.,  {Bildsten} L.,  {Dotter} A.,  {Herwig} F.,  {Lesaffre} P.,   {Timmes} F.,  2011, \mn@doi [\apjs] {10.1088/0067-0049/192/1/3}, 192, 3

\bibitem[\protect\citeauthoryear{{Paxton} et~al.,}{{Paxton} et~al.}{2013}]{Paxton2013}
{Paxton} B.,  et~al., 2013, \mn@doi [\apjs] {10.1088/0067-0049/208/1/4}, 208, 4

\bibitem[\protect\citeauthoryear{{Paxton} et~al.,}{{Paxton} et~al.}{2015}]{Paxton2015}
{Paxton} B.,  et~al., 2015, \mn@doi [\apjs] {10.1088/0067-0049/220/1/15}, 220, 15

\bibitem[\protect\citeauthoryear{{Pearce}, {Tupper}, {Pudritz}  \& {Higgs}}{{Pearce} et~al.}{2018}]{Pearce2018}
{Pearce} B. K.~D.,  {Tupper} A.~S.,  {Pudritz} R.~E.,   {Higgs} P.~G.,  2018, \mn@doi [Astrobiology] {10.1089/ast.2017.1674}, 18, 343

\bibitem[\protect\citeauthoryear{{Peterson}, {Fischer}  \& {LUVOIR Science and Technology Definition Team}}{{Peterson} et~al.}{2017}]{Peterson2017}
{Peterson} B.~M.,  {Fischer} D.,   {LUVOIR Science and Technology Definition Team} 2017, in American Astronomical Society Meeting Abstracts \#229. p. 405.04

\bibitem[\protect\citeauthoryear{Prosdocimi \& {de Farias}}{Prosdocimi \& {de Farias}}{2023}]{Prosdocimi2023}
Prosdocimi F.,  {de Farias} S.~T.,  2023, \mn@doi [Progress in Biophysics and Molecular Biology] {https://doi.org/10.1016/j.pbiomolbio.2023.04.005}, 180-181, 28

\bibitem[\protect\citeauthoryear{{Rampino} \& {Caldeira}}{{Rampino} \& {Caldeira}}{1994}]{Rampino1994}
{Rampino} M.~R.,  {Caldeira} K.,  1994, \mn@doi [\araa] {10.1146/annurev.aa.32.090194.000503}, 32, 83

\bibitem[\protect\citeauthoryear{{Roberge} \& {Moustakas}}{{Roberge} \& {Moustakas}}{2018}]{Roberge2018}
{Roberge} A.,  {Moustakas} L.~A.,  2018, \mn@doi [Nature Astronomy] {10.1038/s41550-018-0543-8}, 2, 605

\bibitem[\protect\citeauthoryear{{Roman}}{{Roman}}{1959}]{Roman1959}
{Roman} N.~G.,  1959, \mn@doi [\aj] {10.1086/108038}, 64, 344

\bibitem[\protect\citeauthoryear{{Salazar}, {McInnes}  \& {Winter}}{{Salazar} et~al.}{2016}]{Salazar2016}
{Salazar} F.~J.~T.,  {McInnes} C.~R.,   {Winter} O.~C.,  2016, \mn@doi [Advances in Space Research] {10.1016/j.asr.2016.04.007}, 58, 17

\bibitem[\protect\citeauthoryear{{Salvador}, {Robinson}, {Fortney}  \& {Marley}}{{Salvador} et~al.}{2024}]{Salvador2024}
{Salvador} A.,  {Robinson} T.~D.,  {Fortney} J.~J.,   {Marley} M.~S.,  2024, \mn@doi [\apjl] {10.3847/2041-8213/ad54c5}, 969, L22

\bibitem[\protect\citeauthoryear{{S{\'a}nchez} \& {McInnes}}{{S{\'a}nchez} \& {McInnes}}{2015}]{Sanchez2015}
{S{\'a}nchez} J.-P.,  {McInnes} C.~R.,  2015, \mn@doi [PLoS ONE] {10.1371/journal.pone.0136648}, 10, e0136648

\bibitem[\protect\citeauthoryear{{Schwieterman}}{{Schwieterman}}{2018}]{Schwieterman2018}
{Schwieterman} E.~W.,  2018, in {Deeg} H.~J.,  {Belmonte} J.~A.,  eds, , Handbook of Exoplanets.
Springer, Cham., p.~69

\bibitem[\protect\citeauthoryear{{Seager}}{{Seager}}{2010}]{Seager2010}
{Seager} S.,  2010, {Exoplanet Atmospheres: Physical Processes}.
Princeton Univ. Press, Princeton, NJ

\bibitem[\protect\citeauthoryear{{Sheikh}, {Huston}, {Fan}, {Wright}, {Beatty}, {Martini}, {Kopparapu}  \& {Frank}}{{Sheikh} et~al.}{2025}]{Sheikh2025}
{Sheikh} S.~Z.,  {Huston} M.~J.,  {Fan} P.,  {Wright} J.~T.,  {Beatty} T.,  {Martini} C.,  {Kopparapu} R.,   {Frank} A.,  2025, \mn@doi [\aj] {10.3847/1538-3881/ada3c7}, 169, 118

\bibitem[\protect\citeauthoryear{Skoglund \& Mustill}{Skoglund \& Mustill}{2025}]{Skoglund_2025}
Skoglund C.,  Mustill A.~J.,  2025, Starshade simulation code: Skoglund and Mustill (2025), \mn@doi{10.5281/zenodo.17175107}, \url {https://doi.org/10.5281/zenodo.17175107}

\bibitem[\protect\citeauthoryear{{Spergel} et~al.,}{{Spergel} et~al.}{2015}]{Spergel2015}
{Spergel} D.,  et~al., 2015, \mn@doi [arXiv e-prints] {10.48550/arXiv.1503.03757}, p. arXiv:1503.03757

\bibitem[\protect\citeauthoryear{{Sugiura}, {Takao}, {Sugihara}, {Sugawara}  \& {Mori}}{{Sugiura} et~al.}{2023}]{Sugiura2023}
{Sugiura} K.,  {Takao} Y.,  {Sugihara} A.~K.,  {Sugawara} Y.,   {Mori} O.,  2023, \mn@doi [Acta Astronautica] {10.1016/j.actaastro.2023.03.040}, 208, 36

\bibitem[\protect\citeauthoryear{{Szapudi}}{{Szapudi}}{2023}]{Szapudi2023}
{Szapudi} I.,  2023, \mn@doi [Proceedings of the National Academy of Science] {10.1073/pnas.2307434120}, 120, e2307434120

\bibitem[\protect\citeauthoryear{{The LUVOIR Team}}{{The LUVOIR Team}}{2019}]{Luvoir2019}
{The LUVOIR Team} 2019, \mn@doi [arXiv e-prints] {10.48550/arXiv.1912.06219}, p. arXiv:1912.06219

\bibitem[\protect\citeauthoryear{{Tremblay} et~al.,}{{Tremblay} et~al.}{2025}]{Tremblay2025}
{Tremblay} C.~D.,  et~al., 2025, \mn@doi [\aj] {10.3847/1538-3881/ad9ea5}, 169, 122

\bibitem[\protect\citeauthoryear{{Tuchow}, {Stark}  \& {Mamajek}}{{Tuchow} et~al.}{2024}]{Tuchow2024}
{Tuchow} N.~W.,  {Stark} C.~C.,   {Mamajek} E.,  2024, \mn@doi [\aj] {10.3847/1538-3881/ad25ec}, 167, 139

\bibitem[\protect\citeauthoryear{{Tusay} et~al.,}{{Tusay} et~al.}{2024}]{Tusay2024}
{Tusay} N.,  et~al., 2024, \mn@doi [\aj] {10.3847/1538-3881/ad823c}, 168, 283

\bibitem[\protect\citeauthoryear{{Vaughan} et~al.,}{{Vaughan} et~al.}{2023}]{Vaughan2023}
{Vaughan} S.~R.,  et~al., 2023, \mn@doi [\mnras] {10.1093/mnras/stad2127}, 524, 5477

\bibitem[\protect\citeauthoryear{Virtanen et~al.,}{Virtanen et~al.}{2020}]{Scipy}
Virtanen P.,  et~al., 2020, \mn@doi [Nature Methods] {10.1038/s41592-019-0686-2}, \href {https://rdcu.be/b08Wh} {17, 261}

\bibitem[\protect\citeauthoryear{Weisstein}{Weisstein}{2024}]{weisstein_circle_intersection}
Weisstein E.~W.,  2024, Circle-Circle Intersection, \url {https://mathworld.wolfram.com/Circle-CircleIntersection.html}

\bibitem[\protect\citeauthoryear{{W}es {M}c{K}inney}{{W}es {M}c{K}inney}{2010}]{Pandas}
{W}es {M}c{K}inney 2010, in {S}t\'efan van~der {W}alt {J}arrod {M}illman eds, {P}roceedings of the 9th {P}ython in {S}cience {C}onference. pp 56 -- 61

\bibitem[\protect\citeauthoryear{{Wordsworth} \& {Kreidberg}}{{Wordsworth} \& {Kreidberg}}{2022}]{Wordsworth2022}
{Wordsworth} R.,  {Kreidberg} L.,  2022, \mn@doi [\araa] {10.1146/annurev-astro-052920-125632}, 60, 159

\bibitem[\protect\citeauthoryear{{Wright}, {Haqq-Misra}, {Frank}, {Kopparapu}, {Lingam}  \& {Sheikh}}{{Wright} et~al.}{2022}]{Wright2022}
{Wright} J.~T.,  {Haqq-Misra} J.,  {Frank} A.,  {Kopparapu} R.,  {Lingam} M.,   {Sheikh} S.~Z.,  2022, \mn@doi [\apjl] {10.3847/2041-8213/ac5824}, 927, L30

\makeatother
\end{thebibliography}




\appendix

\section{Star Selection List}
This appendix consists of a table that displays our results for the starshade simulations of the HWO targets. 

\begin{table*}
	\centering
        \caption{This table shows every star listed in the HWO target list provided by \protect\cite{Harada2024} as well as their tabulated value for metallicity (Met) and distance $d$ from Earth. The mass, age and radius for each star has been fitted using isochrones from MIST. Also stated for each star is our calculated value for the angular separation between the star and an Earth-analogue planet at 1 Gyr as seen from an observer on Earth (stars younger than 1.5 Gyr display no values since we have not performed the simulations for these). The $\left<\Delta_\mathrm{max}\right>$ values are shown for an IWA of 45 and 60 mas respectively (here, empty entries indicates that the planet is always inside the IWA and hence no values could be attained).}
            \label{tab:HWO_stars}
	\begin{tabular}{lccccccccr}
		\hline
        		TIC & HIP & Mass [$M_\odot$] & Age [Gyr] & Met & Rad [$R_\odot$] & d [pc] & Ang. sep 1 Gyr [arcsec] & $\Delta_\mathrm{max}$ 45 mas & $\Delta_\mathrm{max}$ 60 mas \\
		\hline
		TIC 393844873 & HIP 41926 & 0.69 & 19.29 & -0.39 & 0.76 & 12.16 & 0.04 & --- & --- \\
		TIC 417762326 & HIP 42438 & 1.01 & 1.34 & -0.07 & 0.95 & 14.44 & --- & --- & --- \\
		TIC 181273463 & HIP 42808 & 0.82 & 1.92 & 0.01 & 0.75 & 11.19 & 0.05 & 2.28e-10 & --- \\
		TIC 332064670 & HIP 43587 & 0.94 & 7.96 & 0.32 & 0.95 & 12.59 & 0.05 & 7.51e-10 & 5.76e-11 \\
		TIC 62569281 & HIP 43726 & 1.04 & 1.83 & 0.10 & 0.98 & 16.85 & 0.06 & 1.35e-10 & --- \\
		TIC 355127594 & HIP 44897 & 1.08 & 1.67 & 0.04 & 1.04 & 18.95 & 0.06 & 9.00e-11 & --- \\
		TIC 219709102 & HIP 45038 & 1.29 & 3.14 & -0.02 & 1.68 & 20.52 & 0.08 & 1.68e-10 & 1.42e-10 \\
		TIC 8915802 & HIP 47080 & 1.06 & 0.54 & 0.34 & 0.97 & 11.23 & --- & --- & --- \\
		TIC 11310083 & HIP 47592 & 1.08 & 4.41 & -0.06 & 1.23 & 14.95 & 0.08 & 3.97e-10 & 3.33e-10 \\
		TIC 23969522 & HIP 48113 & 1.12 & 6.72 & 0.10 & 1.60 & 18.82 & 0.06 & 6.27e-10 & 2.36e-10 \\
		TIC 172954294 & HIP 49081 & 1.05 & 6.89 & 0.20 & 1.19 & 14.93 & 0.06 & 6.57e-10 & 2.85e-10 \\
		TIC 371520835 & HIP 49908 & 0.66 & 4.23 & 0.21 & 0.64 & 4.87 & 0.06 & 3.20e-09 & 1.15e-09 \\
		TIC 95431211 & HIP 50564 & 1.35 & 2.52 & 0.09 & 1.67 & 21.22 & 0.09 & 1.04e-10 & 8.87e-11 \\
		TIC 453620177 & HIP 50954 & 1.45 & 1.17 & 0.05 & 1.59 & 16.22 & --- & --- & --- \\
		TIC 416519065 & HIP 51459 & 1.06 & 3.28 & -0.12 & 1.11 & 12.95 & 0.09 & 4.33e-10 & 3.79e-10 \\
		TIC 367631379 & HIP 51502 & 1.25 & 1.39 & -0.16 & 1.32 & 22.73 & --- & --- & --- \\
		TIC 447823435 & HIP 51523 & 1.25 & 0.07 & -0.40 & 1.85 & 22.06 & --- & --- & --- \\
		TIC 21535479 & HIP 53721 & 1.01 & 7.31 & 0.02 & 1.21 & 13.89 & 0.07 & 8.53e-10 & 5.92e-10 \\
		TIC 166646191 & HIP 54035 & 0.27 & 0.05 & -0.42 & 0.36 & 2.55 & --- & --- & --- \\
		TIC 57611256 & HIP 56452 & 0.68 & 15.92 & -0.40 & 0.75 & 9.56 & 0.05 & 2.57e-09 & --- \\
		TIC 101641846 & HIP 56997 & 0.89 & 4.64 & -0.05 & 0.86 & 9.58 & 0.07 & 7.68e-10 & 5.95e-10 \\
		TIC 454082369 & HIP 57443 & 0.78 & 16.30 & -0.31 & 0.97 & 9.32 & 0.06 & 4.08e-09 & 2.26e-09 \\
		TIC 366661076 & HIP 57757 & 1.28 & 3.69 & 0.13 & 1.68 & 10.93 & 0.15 & 1.53e-10 & 1.48e-10 \\
		TIC 309599261 & HIP 57939 & 0.59 & 0.06 & -1.33 & 0.60 & 9.17 & --- & --- & --- \\
		TIC 421189312 & HIP 59199 & 1.34 & 1.26 & -0.13 & 1.44 & 14.94 & --- & --- & --- \\
		TIC 1628071 & HIP 61174 & 1.39 & 0.05 & -0.06 & 1.52 & 18.24 & --- & --- & --- \\
		TIC 458445966 & HIP 61317 & 0.91 & 8.43 & -0.20 & 1.03 & 8.47 & 0.09 & 1.22e-09 & 1.13e-09 \\
		TIC 389853353 & HIP 62207 & 0.86 & 0.45 & -0.53 & 1.00 & 17.56 & --- & --- & --- \\
		TIC 445070560 & HIP 64394 & 1.09 & 2.34 & 0.06 & 1.08 & 9.20 & 0.12 & 2.43e-10 & 2.37e-10 \\
		TIC 30293512 & HIP 64408 & 1.25 & 5.01 & 0.16 & 2.12 & 20.46 & 0.07 & 3.88e-10 & 3.04e-10 \\
		TIC 255854921 & HIP 64583 & 1.02 & 6.56 & -0.31 & 1.47 & 18.24 & 0.06 & 7.05e-10 & 3.78e-10 \\
		TIC 373765355 & HIP 64797 & 0.75 & 0.07 & -0.18 & 0.78 & 10.99 & --- & --- & --- \\
		TIC 422478973 & HIP 64924 & 0.88 & 10.76 & -0.01 & 0.98 & 8.53 & 0.08 & 1.63e-09 & 1.43e-09 \\
		TIC 202380743 & HIP 68184 & 0.85 & 0.35 & 0.20 & 0.76 & 10.07 & --- & --- & --- \\
		TIC 83391616 & HIP 69965 & 0.82 & 10.20 & -0.68 & 0.98 & 17.99 & 0.04 & 4.98e-10 & --- \\
		TIC 441709021 & HIP 70497 & 1.22 & 4.11 & -0.03 & 1.70 & 14.53 & 0.11 & 2.84e-10 & 2.81e-10 \\
		TIC 157966796 & HIP 71284 & 1.13 & 2.95 & -0.41 & 1.31 & 15.76 & 0.10 & 2.87e-10 & 2.65e-10 \\
		TIC 471011144 & HIP 71681 & 0.94 & 2.19 & 0.24 & 0.86 & 1.33 & 0.51 & 3.63e-10 & 3.77e-10 \\
		TIC 471011145 & HIP 71683 & 1.08 & 1.52 & 0.20 & 1.23 & 1.33 & 0.74 & 9.30e-10 & 9.16e-10 \\
		TIC 1101124559 & HIP 72659 & 0.70 & 1.26 & 0.14 & 0.65 & 6.75 & --- & --- & --- \\
		TIC 1101124558 & HIP 72659 & 0.86 & 4.20 & -0.14 & 0.82 & 6.75 & 0.10 & 7.17e-10 & 6.27e-10 \\
		TIC 287157634 & HIP 73184 & 0.70 & 7.05 & 0.02 & 0.73 & 5.89 & 0.06 & 5.91e-09 & 3.52e-09 \\
		TIC 229902025 & HIP 73996 & 1.24 & 2.79 & -0.02 & 1.45 & 19.54 & 0.08 & 1.84e-10 & 1.62e-10 \\
		TIC 136916387 & HIP 75181 & 0.79 & 16.49 & -0.34 & 1.05 & 14.74 & 0.04 & 1.35e-09 & --- \\
		TIC 459427073 & HIP 77052 & 0.98 & 3.02 & 0.05 & 0.94 & 14.79 & 0.06 & 3.45e-10 & --- \\
		TIC 296740796 & HIP 77257 & 1.02 & 8.01 & -0.01 & 1.36 & 11.92 & 0.08 & 8.79e-10 & 7.70e-10 \\
		TIC 179348425 & HIP 77358 & 0.97 & 4.92 & 0.10 & 0.96 & 15.25 & 0.05 & 4.65e-10 & --- \\
		TIC 157364190 & HIP 77760 & 1.12 & 0.22 & -0.47 & 1.74 & 15.90 & --- & --- & --- \\
		TIC 377415363 & HIP 78072 & 1.10 & 5.10 & -0.18 & 1.47 & 11.25 & 0.11 & 5.62e-10 & 5.21e-10 \\
		TIC 458494003 & HIP 78459 & 0.90 & 11.81 & -0.22 & 1.33 & 17.51 & 0.04 & 1.05e-09 & 2.40e-10 \\
		TIC 135656809 & HIP 79672 & 0.99 & 5.91 & 0.03 & 1.04 & 14.14 & 0.06 & 5.43e-10 & 2.08e-10 \\
		TIC 350673608 & HIP 80337 & 1.06 & 0.34 & 0.05 & 0.97 & 12.89 & --- & --- & --- \\
		TIC 58092025 & HIP 81300 & 0.87 & 3.60 & 0.03 & 0.82 & 9.89 & 0.06 & 4.72e-10 & 2.54e-10 \\
		TIC 1277142191 & HIP 84405 & 0.78 & 3.53 & -0.22 & 0.72 & 5.95 & 0.09 & 5.92e-10 & 5.14e-10 \\
		TIC 1277142190 & HIP 84405 & 0.78 & 3.20 & -0.22 & 0.73 & 5.95 & 0.09 & 8.99e-10 & 8.39e-10 \\
		TIC 79841001 & HIP 84478 & 0.63 & 5.12 & -0.21 & 0.66 & 5.95 & 0.06 & 5.56e-09 & --- \\
		TIC 217157387 & HIP 84720 & 0.76 & 0.40 & -0.35 & 0.82 & 8.79 & --- & --- & --- \\
		TIC 9728611 & HIP 84862 & 0.79 & 17.21 & -0.39 & 1.16 & 14.59 & 0.04 & 1.61e-09 & --- \\
		\hline
	\end{tabular}
\end{table*}

\begin{table*}
	\centering
	\begin{tabular}{lccccccccr}
		\hline
		TIC & HIP & Mass [$M_\odot$] & Age [Gyr] & Met & Rad [$R_\odot$] & d [pc] & Ang. sep 1 Gyr [arcsec] & $\Delta_\mathrm{max}$ 45 mas & $\Delta_\mathrm{max}$ 60 mas \\
		\hline       
		TIC 75899957 & HIP 84893 & 1.24 & 2.52 & -0.25 & 1.48 & 17.52 & 0.10 & 1.59e-10 & 1.48e-10 \\
		TIC 219880402 & HIP 85235 & 0.71 & 1.86 & -0.46 & 0.76 & 12.79 & 0.04 & --- & --- \\
		TIC 96745915 & HIP 86486 & 1.26 & 2.83 & -0.22 & 1.62 & 20.96 & 0.09 & 1.76e-10 & 1.52e-10 \\
		TIC 238115675 & HIP 86736 & 1.16 & 3.22 & -0.11 & 1.34 & 17.65 & 0.08 & 2.97e-10 & 2.49e-10 \\
		TIC 362661163 & HIP 86796 & 1.13 & 6.45 & 0.29 & 1.38 & 15.60 & 0.07 & 4.88e-10 & 3.56e-10 \\
		TIC 1674663309 & HIP 88601 & 0.87 & 7.68 & 0.06 & 0.87 & 5.11 & 0.12 & 1.19e-09 & 1.13e-09 \\
		TIC 398120047 & HIP 88601 & 0.70 & 4.77 & 0.03 & 0.67 & 5.12 & 0.07 & 2.38e-09 & 1.82e-09 \\
		TIC 329574145 & HIP 88694 & 1.01 & 2.57 & -0.05 & 0.98 & 17.11 & 0.06 & 2.65e-10 & --- \\
		TIC 75946144 & HIP 88972 & 0.73 & 3.51 & -0.18 & 0.79 & 11.10 & 0.04 & --- & --- \\
		TIC 303704858 & HIP 89042 & 1.00 & 7.56 & -0.07 & 1.24 & 17.75 & 0.05 & 5.86e-10 & 1.36e-10 \\
		TIC 233121747 & HIP 89348 & 1.13 & 4.57 & -0.29 & 1.62 & 23.16 & 0.06 & 3.94e-10 & 1.82e-10 \\
		TIC 359981217 & HIP 95447 & 1.08 & 8.52 & 0.38 & 1.39 & 14.92 & 0.06 & 8.05e-10 & 4.26e-10 \\
		TIC 259237827 & HIP 96100 & 0.79 & 8.48 & -0.21 & 0.78 & 5.76 & 0.10 & 1.61e-09 & 1.48e-09 \\
		TIC 58445695 & HIP 97295 & 1.28 & 2.52 & 0.03 & 1.50 & 20.99 & 0.08 & 1.34e-10 & 1.12e-10 \\
		TIC 408842743 & HIP 97675 & 1.24 & 1.09 & 0.12 & 1.49 & 19.49 & --- & --- & --- \\
		TIC 105999792 & HIP 98767 & 0.98 & 10.17 & 0.22 & 1.16 & 16.00 & 0.05 & 7.14e-10 & 9.92e-11 \\
		TIC 352402781 & HIP 98959 & 0.86 & 12.00 & -0.19 & 1.03 & 17.93 & 0.04 & 3.35e-10 & --- \\
		TIC 409891396 & HIP 99240 & 1.04 & 8.16 & 0.35 & 1.20 & 6.10 & 0.14 & 8.99e-10 & 8.67e-10 \\
		TIC 389198736 & HIP 99461 & 0.67 & 0.91 & -0.52 & 0.71 & 6.01 & --- & --- & --- \\
		TIC 326096771 & HIP 99825 & 0.81 & 10.95 & 0.03 & 0.82 & 8.81 & 0.06 & 1.75e-09 & 6.34e-10 \\
		TIC 403585118 & HIP 100017 & 0.99 & 3.85 & -0.10 & 1.00 & 17.48 & 0.05 & 2.78e-10 & --- \\
		TIC 269995013 & HIP 102485 & 1.40 & 0.06 & 0.04 & 1.42 & 14.63 & --- & --- & --- \\
		TIC 29495621 & HIP 103389 & 1.16 & 1.04 & -0.01 & 1.18 & 21.10 & --- & --- & --- \\
		TIC 165602000 & HIP 104214 & 0.65 & 7.24 & -0.13 & 0.64 & 3.50 & 0.10 & 3.31e-09 & 3.12e-09 \\
		TIC 165602023 & HIP 104217 & 0.57 & 4.23 & -0.21 & 0.55 & 3.50 & 0.08 & 2.15e-09 & 1.81e-09 \\
		TIC 159746875 & HIP 105090 & 0.60 & 0.04 & -0.01 & 0.64 & 3.97 & --- & --- & --- \\
		TIC 265488188 & HIP 105858 & 0.80 & 12.76 & -0.70 & 1.09 & 9.26 & 0.08 & 3.10e-09 & 2.61e-09 \\
		TIC 301880196 & HIP 107350 & 1.01 & 2.97 & -0.06 & 1.01 & 18.13 & 0.05 & 2.91e-10 & --- \\
		TIC 147407292 & HIP 107649 & 1.02 & 3.87 & -0.03 & 1.04 & 15.56 & 0.06 & 3.18e-10 & 1.34e-10 \\
		TIC 231698181 & HIP 108870 & 0.69 & 1.47 & -0.13 & 0.73 & 3.64 & --- & --- & --- \\
		TIC 97402436 & HIP 109422 & 1.27 & 2.25 & 0.10 & 1.41 & 18.46 & 0.08 & 1.32e-10 & 1.09e-10 \\
		TIC 259291108 & HIP 110649 & 1.03 & 8.95 & -0.03 & 1.72 & 20.34 & 0.05 & 7.59e-10 & 2.35e-10 \\
		TIC 69889261 & HIP 111449 & 1.36 & 0.12 & 0.03 & 1.42 & 23.02 & --- & --- & --- \\
		TIC 60716322 & HIP 112447 & 1.11 & 5.39 & -0.27 & 1.85 & 16.15 & 0.08 & 7.03e-10 & 6.07e-10 \\
		TIC 206686962 & HIP 113283 & 0.73 & 5.11 & 0.04 & 0.70 & 7.60 & 0.06 & 1.09e-09 & --- \\
		TIC 155315739 & HIP 114046 & 0.41 & 0.09 & -0.22 & 0.47 & 3.29 & --- & --- & --- \\
		TIC 283722336 & HIP 114622 & 0.81 & 0.14 & 0.06 & 0.72 & 6.54 & --- & --- & --- \\
		TIC 24467943 & HIP 114924 & 1.11 & 4.72 & 0.03 & 1.29 & 20.61 & 0.06 & 3.41e-10 & 8.15e-11 \\
		TIC 238432056 & HIP 544 & 0.96 & 1.90 & 0.11 & 0.89 & 13.77 & 0.06 & 2.44e-10 & --- \\
		TIC 289673491 & HIP 910 & 1.00 & 7.25 & -0.36 & 1.51 & 18.89 & 0.06 & 8.60e-10 & 3.74e-10 \\
		TIC 70847587 & HIP 950 & 1.20 & 2.75 & -0.11 & 1.37 & 21.72 & 0.07 & 1.94e-10 & 1.39e-10 \\
		TIC 425935521 & HIP 1599 & 0.92 & 8.09 & -0.21 & 1.06 & 8.61 & 0.10 & 1.11e-09 & 1.01e-09 \\
		TIC 267211065 & HIP 2021 & 1.07 & 7.08 & -0.12 & 1.84 & 7.46 & 0.15 & 9.60e-10 & 9.23e-10 \\
		TIC 434210589 & HIP 3093 & 0.85 & 12.44 & 0.14 & 0.90 & 11.11 & 0.05 & 1.58e-09 & --- \\
		TIC 80431620 & HIP 3583 & 1.04 & 0.04 & -0.04 & 0.92 & 15.05 & --- & --- & --- \\
		TIC 257393898 & HIP 3765 & 0.68 & 19.92 & -0.26 & 0.73 & 7.44 & 0.05 & 5.63e-09 & 6.70e-10 \\
		TIC 445258206 & HIP 3821 & 0.88 & 9.49 & -0.28 & 1.07 & 5.01 & 0.15 & 1.87e-09 & 1.85e-09 \\
		TIC 3962869 & HIP 3909 & 1.07 & 3.63 & -0.12 & 1.15 & 15.92 & 0.07 & 2.86e-10 & 2.00e-10 \\
		TIC 285544488 & HIP 4151 & 1.21 & 4.57 & 0.05 & 1.66 & 18.80 & 0.08 & 2.96e-10 & 2.46e-10 \\
		TIC 229092427 & HIP 5862 & 1.21 & 0.40 & 0.16 & 1.26 & 15.26 & --- & --- & --- \\
		TIC 52194638 & HIP 5896 & 1.30 & 2.65 & 0.03 & 1.59 & 23.26 & 0.08 & 1.29e-10 & 1.02e-10 \\
		TIC 189576919 & HIP 7513 & 1.25 & 3.80 & 0.08 & 1.61 & 13.49 & 0.11 & 1.70e-10 & 1.67e-10 \\
		TIC 231005052 & HIP 7751 & 0.79 & 0.08 & -0.19 & 0.70 & 8.20 & --- & --- & --- \\
		TIC 231005905 & HIP 7751 & 0.72 & 3.04 & -0.24 & 0.78 & 8.19 & 0.06 & 4.41e-09 & --- \\
		TIC 229137615 & HIP 7978 & 1.10 & 1.91 & -0.04 & 1.10 & 17.35 & 0.07 & 1.47e-10 & 9.61e-11 \\
		TIC 113710966 & HIP 7981 & 0.81 & 10.53 & -0.04 & 0.83 & 7.64 & 0.07 & 1.82e-09 & 1.45e-09 \\
		TIC 419015728 & HIP 8102 & 0.72 & 3.84 & -0.51 & 0.83 & 3.65 & 0.14 & 6.06e-09 & 5.72e-09 \\
		TIC 373694425 & HIP 8362 & 0.90 & 2.66 & 0.03 & 0.83 & 10.04 & 0.07 & 5.15e-10 & 3.36e-10 \\
		TIC 72748794 & HIP 10798 & 0.72 & 13.79 & -0.49 & 0.76 & 12.83 & 0.04 & 4.21e-10 & --- \\
		TIC 166853853 & HIP 12653 & 1.22 & 0.22 & 0.15 & 1.16 & 17.36 & --- & --- & --- \\
		TIC 302158903 & HIP 12777 & 1.16 & 2.92 & 0.01 & 1.27 & 11.15 & 0.12 & 2.86e-10 & 2.79e-10 \\
		TIC 326242565 & HIP 12843 & 1.23 & 2.30 & 0.07 & 1.35 & 14.28 & 0.10 & 1.71e-10 & 1.56e-10 \\
		\hline
	\end{tabular}
\end{table*}

\begin{table*}
	\centering
	\begin{tabular}{lccccccccr}
		\hline
		TIC & HIP & Mass [$M_\odot$] & Age [Gyr] & Met & Rad [$R_\odot$] & d [pc] & Ang. sep 1 Gyr [arcsec] & $\Delta_\mathrm{max}$ 45 mas & $\Delta_\mathrm{max}$ 60 mas \\
		\hline
		TIC 30016911 & HIP 13402 & 0.89 & 0.06 & 0.07 & 0.78 & 10.36 & --- & --- & --- \\
		TIC 116988032 & HIP 14632 & 1.13 & 5.62 & 0.09 & 1.40 & 10.58 & 0.11 & 3.79e-10 & 3.61e-10 \\
		TIC 88523071 & HIP 14879 & 1.11 & 5.42 & -0.22 & 1.79 & 14.00 & 0.10 & 6.80e-10 & 6.60e-10 \\
		TIC 279649057 & HIP 15330 & 0.85 & 8.64 & -0.23 & 0.91 & 12.04 & 0.06 & 1.14e-09 & 3.73e-10 \\
		TIC 279649049 & HIP 15371 & 0.89 & 8.24 & -0.23 & 0.98 & 12.04 & 0.06 & 1.02e-09 & 6.38e-10 \\
		TIC 343813545 & HIP 15457 & 0.99 & 3.02 & 0.04 & 0.95 & 9.28 & 0.09 & 4.46e-10 & 4.14e-10 \\
		TIC 301051051 & HIP 15510 & 0.76 & 5.38 & -0.39 & 0.91 & 6.04 & 0.09 & 5.43e-09 & 5.06e-09 \\
		TIC 262843771 & HIP 16245 & 1.30 & 2.53 & -0.17 & 1.66 & 21.78 & 0.09 & 1.17e-10 & 1.13e-10 \\
		TIC 118572803 & HIP 16537 & 0.81 & 0.56 & -0.08 & 0.75 & 3.22 & --- & --- & --- \\
		TIC 311092847 & HIP 16852 & 1.24 & 0.07 & -0.08 & 1.62 & 13.92 & --- & --- & --- \\
		TIC 38511251 & HIP 17378 & 1.20 & 2.55 & 0.09 & 2.24 & 9.09 & 0.15 & 4.78e-10 & 4.81e-10 \\
		TIC 121078878 & HIP 17651 & 1.43 & 0.39 & 0.08 & 1.68 & 17.78 & --- & --- & --- \\
		TIC 9150015 & HIP 18859 & 1.24 & 0.08 & 0.10 & 1.21 & 18.71 & --- & --- & --- \\
		TIC 353257675 & HIP 19335 & 1.31 & 0.06 & 0.19 & 1.24 & 21.19 & --- & --- & --- \\
		TIC 67772871 & HIP 19849 & 0.75 & 1.02 & -0.29 & 0.81 & 5.01 & --- & --- & --- \\
		TIC 117979951 & HIP 22263 & 1.01 & 1.89 & 0.00 & 0.96 & 13.24 & 0.07 & 1.85e-10 & 1.41e-10 \\
		TIC 399665349 & HIP 22449 & 1.25 & 1.04 & 0.03 & 1.32 & 8.07 & --- & --- & --- \\
		TIC 213041474 & HIP 23311 & 0.85 & 2.11 & 0.31 & 0.78 & 8.84 & 0.06 & 4.06e-10 & --- \\
		TIC 381949122 & HIP 23693 & 1.03 & 3.21 & -0.18 & 1.06 & 11.69 & 0.09 & 4.50e-10 & 3.97e-10 \\
		TIC 27136704 & HIP 23835 & 1.30 & 0.01 & -0.20 & 1.59 & 15.92 & --- & --- & --- \\
		TIC 409104974 & HIP 24813 & 1.03 & 7.66 & 0.06 & 1.28 & 12.56 & 0.08 & 7.42e-10 & 6.29e-10 \\
		TIC 142103211 & HIP 25110 & 1.29 & 2.51 & 0.12 & 1.50 & 20.79 & 0.08 & 1.33e-10 & 1.10e-10 \\
		TIC 47346402 & HIP 25278 & 1.11 & 3.33 & 0.01 & 1.18 & 14.58 & 0.08 & 3.51e-10 & 3.28e-10 \\
		TIC 261136679 & HIP 26394 & 1.09 & 3.91 & 0.09 & 1.15 & 18.29 & 0.06 & 2.64e-10 & 6.64e-11 \\
		TIC 311063391 & HIP 26779 & 0.87 & 6.21 & 0.10 & 0.84 & 12.27 & 0.05 & 5.49e-10 & --- \\
		TIC 93279196 & HIP 27072 & 0.75 & 5.68 & -0.14 & 0.72 & 8.89 & 0.05 & 1.02e-09 & --- \\
		TIC 93280676 & HIP 27072 & 1.15 & 0.24 & -0.08 & 1.28 & 8.90 & --- & --- & --- \\
		TIC 176521059 & HIP 27435 & 0.86 & 8.48 & -0.22 & 0.92 & 15.21 & 0.05 & 6.10e-10 & --- \\
		TIC 141810080 & HIP 29271 & 0.95 & 7.06 & 0.10 & 0.99 & 10.21 & 0.07 & 9.20e-10 & 7.57e-10 \\
		TIC 46125330 & HIP 29650 & 1.32 & 0.06 & 0.03 & 1.34 & 21.78 & --- & --- & --- \\
		TIC 437886584 & HIP 29800 & 1.24 & 0.71 & -0.03 & 1.36 & 19.59 & --- & --- & --- \\
		TIC 141523112 & HIP 32439 & 1.08 & 3.67 & -0.10 & 1.17 & 18.20 & 0.06 & 2.53e-10 & 1.29e-10 \\
		TIC 307754027 & HIP 32480 & 1.13 & 3.59 & 0.11 & 1.22 & 16.61 & 0.07 & 2.20e-10 & 1.65e-10 \\
		TIC 282210766 & HIP 32984 & 0.77 & 0.07 & 0.02 & 0.69 & 8.74 & --- & --- & --- \\
		TIC 80226651 & HIP 33277 & 0.96 & 6.91 & -0.12 & 1.08 & 17.40 & 0.05 & 5.91e-10 & --- \\
		TIC 130645536 & HIP 34065 & 0.87 & 12.73 & -0.22 & 1.21 & 17.06 & 0.04 & 8.34e-10 & --- \\
		TIC 156890613 & HIP 35136 & 0.87 & 11.66 & -0.32 & 1.16 & 16.85 & 0.04 & 1.00e-09 & 5.64e-11 \\
		TIC 328324648 & HIP 36439 & 1.05 & 5.03 & -0.28 & 1.29 & 20.40 & 0.06 & 4.67e-10 & 1.80e-10 \\
		TIC 150796339 & HIP 38423 & 1.21 & 1.40 & -0.08 & 1.25 & 18.33 & --- & --- & --- \\
		TIC 372914091 & HIP 38908 & 0.90 & 8.24 & -0.31 & 1.05 & 16.17 & 0.05 & 7.94e-10 & 1.55e-10 \\
		TIC 307624961 & HIP 40693 & 0.86 & 8.57 & -0.03 & 0.88 & 12.58 & 0.05 & 8.81e-10 & --- \\
		TIC 302188141 & HIP 40843 & 1.02 & 5.97 & -0.28 & 1.32 & 18.22 & 0.06 & 5.43e-10 & 2.36e-10 \\
		TIC 234968549 & HIP 114948 & 1.15 & 0.78 & -0.03 & 1.16 & 20.44 & --- & --- & --- \\
		TIC 419919445 & HIP 116771 & 1.11 & 5.57 & -0.14 & 1.59 & 13.71 & 0.09 & 5.06e-10 & 4.74e-10 \\
		\hline
	\end{tabular}
\end{table*}


\bsp	
\label{lastpage}
\end{document}